%% file: main.tex
\RequirePackage{fix-cm}
\PassOptionsToPackage{table}{xcolor}
\documentclass[smallextended]{svjour3}
\smartqed  
\usepackage{booktabs}
\usepackage{xspace}
\usepackage{tabularx}
\usepackage{ulem}
\usepackage{threeparttable}
\usepackage{rotating}
\usepackage{graphicx}
\usepackage{lscape}
\usepackage{amsmath}
\usepackage{soul}
\usepackage{url}
\usepackage[linesnumbered,ruled,lined]{algorithm2e} 
\usepackage{float}
\usepackage{multirow}
\usepackage{diagbox}
\usepackage{makecell}
\usepackage{booktabs}
\usepackage{tabularx}
\usepackage{subcaption}
\usepackage{enumitem}
\usepackage{longtable}
\usepackage{subcaption}
\usepackage{tabularx}
\usepackage{lipsum} 
\usepackage{booktabs}
\usepackage{multirow}
\usepackage{cite}
\usepackage{caption}
\usepackage{subcaption} 
\usepackage{booktabs}   
\usepackage{lipsum}  
 \usepackage{mathptmx}      
\usepackage{hyperref}
\usepackage{url}  

\usepackage{booktabs}
\usepackage{siunitx}
\usepackage{tabularx}
\usepackage{threeparttable}
\usepackage{tcolorbox}
\tcbuselibrary{skins,breakable}
\usepackage{xcolor}
\DeclareUnicodeCharacter{00B1}{\ensuremath{\pm}}
\DeclareUnicodeCharacter{03C0}{\ensuremath{\pi}}
\DeclareUnicodeCharacter{2081}{\textsubscript{1}}

\newcommand{\etal}{et al.}
\newcommand{\Tbl}{Table }
\newcommand{\totalNumFails}{77,354 }
\newcommand{\Fig}{Fig. }
\newcommand{\totalNumObj}{10,316 }
\newcommand{\totalFeaNum}{33 }
\newcommand{\ranSampNum}{371}

\definecolor{goldenred}{RGB}{150,84,84}
\definecolor{customblue}{RGB}{70,130,180}
\newcommand{\highlight}[1]{\textbf{\textcolor{goldenred}{#1}}}

\newcommand{\ie}{i.e., }
\newcommand{\eg}{e.g., }

\renewcommand{\arraystretch}{1.2}
\newcounter{featureblockctr}
\newtcolorbox{featureblock}[4][]{
  colback=gray!5!white, colframe=gray!50!black,
  title=\textbf{#2}, fonttitle=\bfseries,
  enhanced jigsaw, breakable,
  top=1mm, bottom=1mm, left=1mm, right=1mm,
  before upper={\refstepcounter{featureblockctr}\label{#4}},
  after upper={\par\small\textit{\thefeatureblockctr #3}},
  #1
}

\begin{document}
\begin{sloppypar}
\title{Is this Build Failure Related to my Patch? An Empirical Study of Unrelated Build Failures in Continuous Integration
}

\author{Yonghui (Andie) Huang · Daniel Alencar da Costa · Grant Dick · Mariam El Mezouar}

\institute{Andie Huang \at
              University of Otago
              \\
              School of Computing \\
              \email{andie.huang@postgrad.otago.ac.nz}           
           \and
           Daniel Alencar da Costa \at
           University of Otago
              \\
           School of Computing
           \\
           \email{danielcalencar@otago.ac.nz} 
           \and
           Grant Dick \at
           University of Otago
              \\
           School of Computing
           \\
           \email{grant.dick@otago.ac.nz}
           \and
           Mariam El Mezouar \at
           Royal Military College of Canada
           \\
           Department of Mathematics and Computer Science
           \\
           \email{mariam.el-mezouar@rmc.ca}
}

\maketitle

\begin{abstract}
\input{src/abstract.tex}
\end{abstract}

\section{Introduction}
\label{intro}
\input{src/introduction.tex}

\section{The Problem of Unrelated Build Failures}
\label{motivating:example}
\input{src/motivating_example}

\section{Research Methodology}
\label{methodology}

\input{src/Methodology.tex}

\section{Results}
\label{results}
\input{src/results.tex}

\section{Related work}
\label{related:work}
\input{src/relatedwork}

\section{Discussion}
\label{sec:discussion}
\input{src/discussion}

\section{Threats to Validity}
\label{threasToValidity}
\input{src/threatsToValidity.tex}

\section{Conclusion}
\label{conclusion}
\input{src/conclusion.tex}

\bibliographystyle{acm}
\bibliography{ref}
\section{Appendix}
\input{src/appendix/appendix_2}
\end{sloppypar}
\end{document}

%% file: src/abstract.tex
In a hectic Continuous Integration (CI) environment, where several builds are triggered concurrently, legitimate build failures (e.g., not caused by flaky tests) may not always be related to the current push. 
These unrelated build failures can burden developers as they devote hours to attest whether errors are truly associated with their present changes. 
In this paper, we extract \totalNumFails CI build failures from 7 open source projects to understand and identify unrelated build failures.
We attempt to provide an indication for developers about whether a build failure is likely to be related to the current push or not.
Our results reveal that developers likely invest a median of 4 hours to determine whether a build failure is (un)related to their pushes.
We perform a document analysis on a sample of \ranSampNum~unrelated build failures (based on the 95\% confidence level and 5\% confidence interval from \totalNumObj potentially unrelated failures) to understand why build failures are deemed as unrelated by developers.
The themes generated from our document analysis reveal that \textit{unrelated tests failures} represent 20\% of the cases of why build failures are deemed unrelated by developers.
To predict whether a build failure is unrelated to the current push, we extract \totalFeaNum features from issue reports, issue comments, and from the commits pertaining to the triggering push. 
We build semi-supervised PU-learning models over seven Apache projects and achieve precision ranging from $0.70 \pm 0.01$ to $0.88 \pm 0.02$
, recall ranging from $0.30 \pm 0.03$ to $1.00 \pm 0.00$, and F1-scores ranging from $0.44 \pm 0.03$ to $0.91 \pm 0.00$, while the area under the ROC curve (AUC) spans $0.63 \pm 0.02$ to $0.97 \pm 0.03$.
Our analysis of feature importance reveals that (i) the time taken from a submitted patch to the build-triggering push (CI latency), (ii) build failures sharing similar error messages with recent failures, and (iii) the number of comments preceding the build failure, are all efficient indicators for identifying potential unrelated build failures. The semi-supervised approach proposed in this work can help developers identify build failures that are unrelated to their current push, helping their decision-making process.

%% file: src/introduction.tex
Continuous Integration (CI), a software development practice that involves the frequent integration of code changes, has gained traction in both industry and academia~\cite{fowler2006continuous,duvall2007continuous,shahin2017continuous,donca2022method}. Researchers have thoroughly investigated the potential benefits and costs of CI in the last years~\cite{soares2022effects}. For instance, CI has empirically shown associations with increased productivity~\cite{bernardo2018studying,vasilescu2015quality,santos2022investigating} despite potential challenges such as extra efforts required to address CI configuration issues ~\cite{zampetti2019study}.

According to Duvall et al.~\cite{duvall2007continuous}, CI is a combination of several interrelated practices, including frequent check-ins (e.g., frequent \textit{Git Pushes}), short build cycles, a comprehensive test suite, and proper build feedback (or  \textit{bells and whistles}).
Regarding \textit{proper build feedback}, it is important that development teams have visibility into when builds fail, who caused the failure, and the reason for failure. Indeed, Duvall et al.~\cite{duvall2007continuous} emphasize the utmost importance of fixing a build failure before moving on to other tasks. This is because developers may relax their adherence to CI practices if they perceive that builds are constantly failing. For example, developers may choose to forgo thorough checks of their changes before pushing to the mainline (i.e., the build is always failing anyway). 

Notwithstanding the importance of fixing build failures, given the prevalence of distributed software development settings, where multiple pushes are made concurrently, it becomes increasingly challenging to fix build failures immediately. In fact, Ghaleb et al.~\cite{ghaleb2019studying} revealed a phenomenon called \textit{cascading build breakages}, where a build fails not due to the changes of the current push, but due to changes inherited from a previous push that initially caused the build to fail. As such, an adverse effect of \textit{cascading build breakages} is that developers often wonder whether their code changes are directly responsible for the build failure or if the failure is unrelated to their changes, possibly caused by another developer's push.

Much research has studied the challenges involved in adopting CI, including complexities around build notifications~\cite{elazhary2021uncovering,shahin2017continuous,zampetti2020empirical}. Elazhary et al.~\cite{elazhary2021uncovering} studied three companies using CI and reported that they face ``notification fatigue'' due to frequent builds, while also requiring discipline to monitor build status. Zampetti et al.~\cite{zampetti2020empirical} studied CI practitioners to uncover bad patterns related to CI usage. Regarding the status of builds, the authors found that build notifications can often be verbose, containing unrelated information, while also containing the output of different build tasks, which can confuse developers when it comes to fixing build failures. 

While previous research has highlighted issues related to sub-optimal build notifications, our work is the first to address the problem of automatically identifying unrelated build failures. We define an unrelated build failure as a failure that occurs during the continuous integration (CI) process but is not caused by the current code push. 
We do not define “unrelated failures” as failures that are entirely unrelated to the codebase or project as a whole. Rather, we refer to failures that are unrelated to the most recent code push. 
More specifically, an “\textbf{unrelated build failure}” as one where (i) no error messages, stack traces, or test failures can be traced back to the source files modified in the associated code push, (ii) the root cause lies outside the submitted code changes before the merging stage, or (iii) follow-up developer comments explicitly confirm that the issue is unrelated to the submitted code.
In practice, unrelated failures can lead to significant developer frustration and wasted effort, as teams may spend time investigating issues that are not actionable for their current change. If developers can be certain that a build failure is not related to their push, they can use this information to make decisions, such as proceeding with the integration or better debugging the build breakage. This is particularly crucial in distributed development settings, where different collaborators
simultaneously contribute to the project, such as open-source projects or remotely
distributed teams~\cite{lanubile2010collaboration,hertel2003motivation}.

It is important to distinguish unrelated build failures from flaky tests.
\textit{Flaky tests}\cite{luo2014empirical} refer to automated tests that exhibit non-deterministic behavior, producing inconsistent results despite unchanged code. 
Unrelated build failures, on the other hand, pertain to legitimate failures in the build process caused by changes unrelated to the current code push.
For example, configuration issues or unforeseen problems due to differences in environments, can contribute to build failures related to code pushes. 
As such, our goal in this study is to understand and potentially predict when unrelated build failures occur. We conduct seven project-level experiments of Apache projects, where developers often comment on the validity of build failures, exerting efforts to verify whether these failures are unrelated to the current push. 

To scale our study, we employ a keyword-detection approach to identify an initial set of \textit{potential} \totalNumObj unrelated build failures.
We then use the initial set of potentially unrelated build failures to manually curate a representative sample of \ranSampNum (based on 95\% confidence level and 5\% confidence interval out of \totalNumObj samples) truly unrelated build failures. We perform a thematic analysis of this sample to understand why developers deem build failures as unrelated.

Once we understand the reasons why build failures are deemed as unrelated, we use this knowledge to extract features that can be used to predict unrelated build failures. Predicting unrelated build failures can help developers better understand their build notification. For example, the build notification can indicate that errors in a build are potentially unrelated to the developer's triggering push with an accompanying  probability. 
Although it is important to maintain successful build statuses~\cite{duvall2007continuous}, developers could make more informed decisions regarding how to proceed with the merging of contributions when an unrelated failure is present. Moreover, knowing that a build failure is unrelated to the current push may help the debugging process of the build failure as developers would know that the build failure is unrelated to the scope of the changes within the current push. 

We use a semi-supervised approach known as \textit{Positive and Unlabeled} (PU) learning~\cite{liu2002partially,li2003learning,yu2002pebl,nguyen2011positive,yang2012positive,li2010positive} to predict unrelated build failures. We develop PU-learning models (including the classic Elkan \& Noto and weighted Elkan \& Noto models) for each of our seven project-level experiments. 
Our semi-supervised models demonstrate robust performance, with median AUC values ranging from 0.62 to 0.97 across the seven projects analyzed, reflecting how the datasets varied in complexity and patterns, creating different challenges for prediction.
Moreover, the models perform better than random guessing and can provide a reasonable indication as to whether a build failure is related to the current code push.
We then conduct a permutation feature importance analysis to identify the most influential features in our PU-learning models, following the methodologies established in previous studies~\cite{rajbahadur2017impact, tantithamthavorn2018impact, jiarpakdee2020impact}.
Our findings and prediction model offer actionable guidance for developers when a build failure is predicted to be unrelated to recent code changes. By knowing that a certain failure is unrelated to the current push, we provide developers with a deeper understanding of the scope of a given build failure. Rather than being ignored, such failures should be triaged differently from code-related issues. For example, by re-running the build to confirm whether the failure persists or by inspecting infrastructure logs to identify issues in the build environment or external dependencies. This helps avoid unnecessary debugging of recent code. In addition to assisting with short-term triage, repeated unrelated failures may reveal misconfiguration of CI, infrastructure network issues, and resource dependencies. Therefore, teams should systematically monitor CI outcomes. Notifications of unrelated build failures help developers decide whether to investigate the code further or simply re-run the build. This is consistent with recent recommendations by Santos~\etal~\cite{santos2025need}, who emphasize the need to monitor CI practices to enhance both reliability and development efficiency.
In the next section, we explore the challenge of \textit{unrelated build failures}.

%% file: src/motivating_example.tex
\begin{figure}[]
\centering
\includegraphics[width=0.7\textwidth, height=6.5cm]{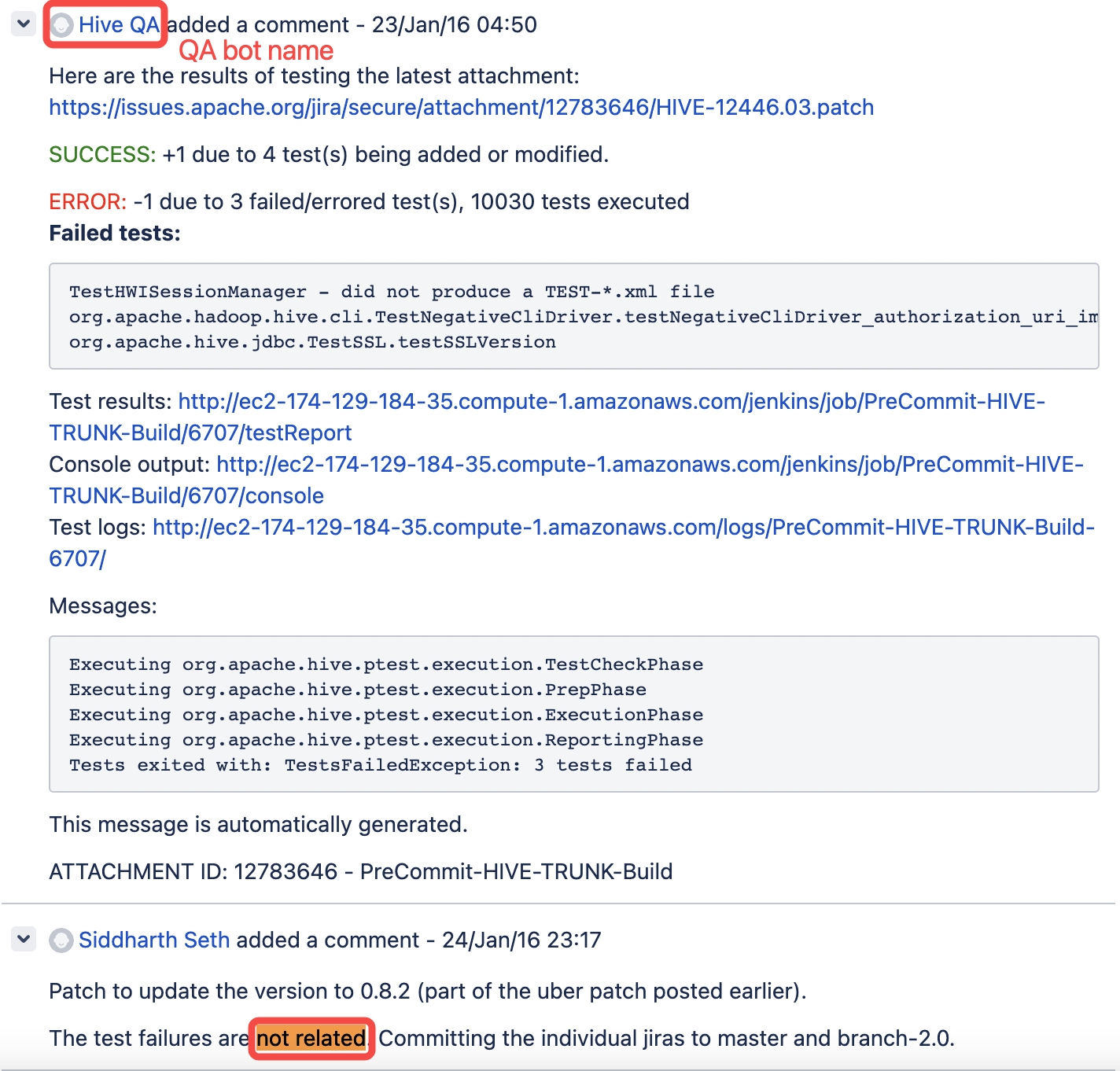}
    \caption{An example of a CI bot comment in a historical issue report after the completion of the CI build}
    \label{method:cibot_ex}
\end{figure}

An unrelated build failure occurs when a build fails due to reasons that are unrelated to the current triggering push. Different from flaky tests~\cite{bell2018deflaker,luo2014empirical,eloussi2015determining,terragni2020container,palomba2017notice,silva2020shake,thorve2018empirical,eck2019understanding}, unrelated failures are legitimate (as opposed to being intermittent) but are unrelated with the scope of changes of the current push. One could argue that unrelated build failures would not occur if CI practices are properly adopted, e.g., code should only be merged upon passing builds. Nevertheless, in projects where multiple developers work concurrently, as in our case and in the projects studied by Ghaleb \etal~\cite{ghaleb2019empirical}, one developer’s change may inadvertently affect another’s work. This can lead to unexpected build failures that are unrelated to the latter’s contributions. 
Additionally, as highlighted by Felidré \etal~\cite{felidre2019continuous}, projects using CI practices do not always adhere to the highest standards of CI practices, which can potentially cause an increased occurrence of unrelated build failures. As such, our research also acknowledges that achieving perfection in adhering to CI practices may not always be feasible in real-world scenarios.

For example, \Fig \ref{method:cibot_ex} depicts an unrelated build failure that occurred in the HIVE project (issue report HIVE-12446\footnote{\url{https://issues.apache.org/jira/browse/HIVE-12446}}).
On January 23, 2016, at 04:50, a build failure prompted an automatic comment posted by the QA bot. Approximately 19 hours later, developers confirmed the purpose of the push, which was to update the \texttt{Tez} version to 0.8.2 in both the \texttt{master} and \texttt{branch-2.0} branches. They then verified that the test failures were unrelated to the push and stated their intention to commit the modifications to the specified branches. Had developers known from the outset that the build failure was unrelated to their push, they could have saved time and effort by making their decision earlier in the development process. 

\begin{figure}
    \centering
    \includegraphics[width=1.0\textwidth, height=5cm]{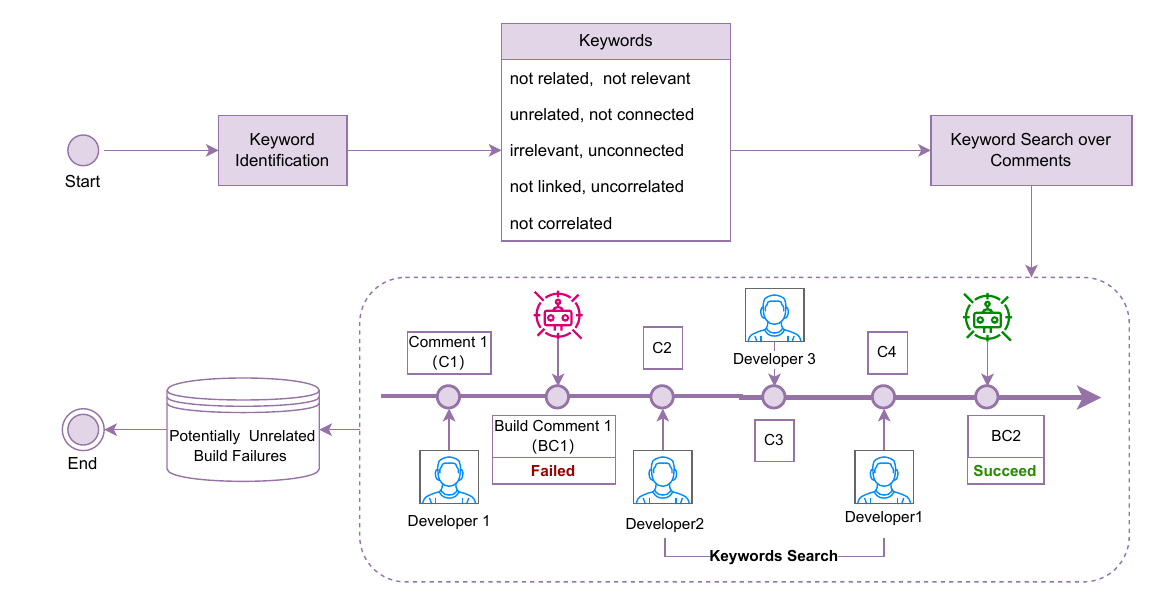}
    \caption{Approach of Heuristic-based Labeling (HL)}
    \label{fig:HL_based_fig}
\end{figure}

We manually inspected a random sample of 383 build failures drawn from a total of 77,354 failure records across the seven studied projects, encompassing both related and unrelated failures, determined using a 95\% confidence level and 5\% confidence interval. Among the 383 randomly selected failures, we found 45 unrelated failures explicitly marked by developers using keywords such as ``unrelated'' and similar terms.  This inspection confirmed that developers do indeed use terms like ``unrelated'', ``not relevant'', ``not connected'', and similar language to describe unrelated failures in practice across our studied seven projects. 
This supports the validity of our keyword-based strategy for identifying such cases. 
However, we acknowledge that keyword usage may vary across projects and that not all unrelated failures are explicitly labeled. To address this, we distinguish between two types of unrelated failures: (1) those containing clear keywords that indicate the failure is not related to a code push, and (2) those without explicit indicators, leaving their cause ambiguous. The former can be identified through a Heuristic-based Labeling (HL) approach (explained below), whereas the latter are more challenging. Since failures not captured by keyword matches cannot be confidently regarded as unrelated, the setting is better suited to a semi-supervised learning paradigm. Therefore, we adopt a Positive–Unlabeled (PU) learning framework: failures with relevant keywords are treated as positive examples, while all others remain unlabeled, acknowledging that the absence of a keyword does not imply relevance to a code push. This design allows the model to benefit from weak supervision while remaining robust to inconsistent or incomplete developer annotations.

\textbf{Heuristic-based Labeling (HL)}. To increase the breadth of our understanding regarding unrelated build failures, we begin by adopting a naïve approach based on keywords to automatically identify build failures that are \textbf{potentially} unrelated. We call this approach {\textit{ Heuristic-based Labeling}}, which is shown in 
\Fig\ref{fig:HL_based_fig}.
Firstly, we create our list of keywords based on a sample of comments in which developers indicated that the build failure was not linked to their push. 
The set of keywords we use is: {\textit{not related}, \textit{unrelated}, \textit{irrelevant}, \textit{not relevant}, \textit{not connected}, \textit{unconnected}, \textit{not linked}, \textit{uncorrelated}, \textit{not correlated}}.
Next, we iterate over the historical comments in an issue report. Once there is a build failure comment (\eg BC1 in \Fig\ref{fig:HL_based_fig}), we analyze the following comments (\eg C2 to C4 in \Fig\ref{fig:HL_based_fig}) until the next build comment (\eg BC2 in \Fig\ref{fig:HL_based_fig}). If we identify the keywords listed above within the set of analyzed comments by developers, we flag the build failure (\eg BC1 in \Fig\ref{fig:HL_based_fig}) as \textbf{potentially} unrelated.

\begin{figure}[]
   \includegraphics[width=0.9\textwidth]
   {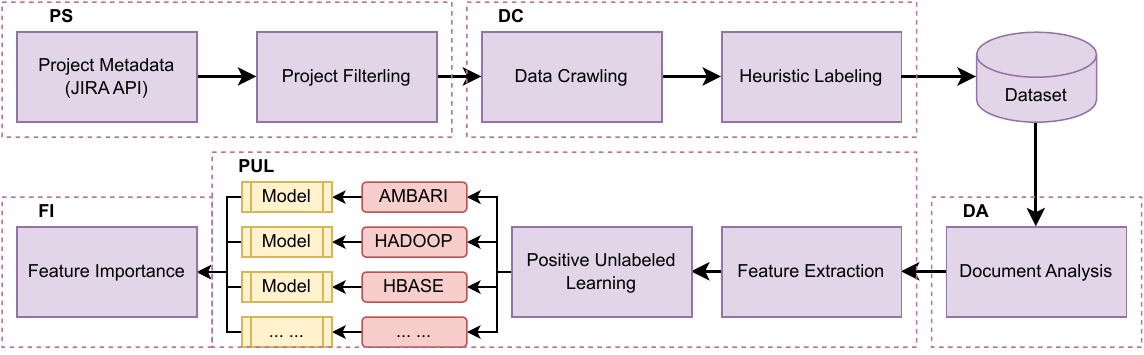}
   \caption{The overview of the process in our study.}
   \label{method:fig1}
\end{figure}

\Fig \ref{method:fig1} shows an overview of our research process, which illustrates our end-to-end research process comprising five interconnected phases: 1) Project Selection (PS) identifies Apache projects with mature CI ecosystems. 2) Data Collection (DC) obtains build failures, issue reports, and developer comments.3) Document Analysis (DA) examines 371 failures to derive why developers deem failures unrelated. 4) PU-Learning (PUL) trains semi-supervised models using features informed by (3). Feature Importance (FI) identifies key predictors of unrelated failures. We now detail these phases.

\textbf{Project Selection (PS).} The first step is concerned with the project selection process. For the purpose of studying CI projects, we have chosen ASF (Apache Software Foundation) projects as the CI culture has matured within the ASF since the early 2000s. In fact, the ASF is the birthplace of the first-ever Java CI tool--\textsc{Apache Gump}, which represented the nightly build of the entire Java stack.\footnote{\url{https://gump.apache.org}} ASF is also where \textsc{Ant} was developed, and its XML output `\texttt{<junit>}' has become the standard format for \textsc{Junit} results. Furthermore, the \textsc{Hadoop} project introduced \textsc{Apache Yetus}\footnote{\url{https://yetus.apache.org}}, a tool designed to be launched whenever a pull request is made. \textsc{Yetus} runs tests, prints the results, and provides code veto.\footnote{\url{https://issues.apache.org/jira/browse/HADOOP-15281}} For instance, a push that has not been tested by \textsc{Yetus} would not proceed to code reviews. Considering the array of developments within the ASF regarding CI tooling and support for CI practices, we believe that ASF projects are suitable for our work. 
 
\input{src/tables/method_tbl1.tex}
 
Given that the ASF uses JIRA to manage its projects, we employed the JIRA API\footnote{https://pypi.org/project/jira/} to extract the necessary metadata, such as issue reports and the main programming languages of all ASF open-source projects that are available (as shown in \Tbl \ref{table:project_metadata_ci_cd}). 
ASF projects using a CI pipeline will normally employ a CI bot that generates automatic comments representing the build logs and statuses whenever a CI build is completed (see Fig.~\ref{method:cibot_ex}). We use these automatic comments to extract the CI build information of our ASF projects. For example, in the \textsc{HIVE} project, we collect all comments from the `Hive QA' author, which represents the build runs of said project. Given that different ASF projects assign different names for their QA bots, we collect the history of authors involved in the discussion comments of each project, searching for the `QA' keyword. Typically the QA bot will follow the `\{PROJECT NAME\} QA' pattern.
 
\Tbl \ref{table:project_metadata_ci_cd} shows the details of metadata we collect through JIRA API.
The table shows the top 20 projects out of the 100 projects' metadata we collected. 
The metadata for the remaining 80 projects can be found in our replication package for reference\cite{githubrepo}.
For each of the 100 projects, we check the total number of issue reports and whether the project contains CI build logs in their comments history. 
We arrange the projects in descending order in terms of the number of issue reports, since we seek to select projects with a larger amount of data as possible for further analyses.
We restrict our focus to projects predominantly programmed in Java and exclude those projects lacking CI build logs on their issues.
We also exclude projects that contain a limited number of CI build logs (\ie less than 10\%) since we wish to focus on the more active projects. 
As a result, from a total of 100 projects, we choose \textsc{AMBARI}, \textsc{HBASE}, \textsc{HIVE}, \textsc{HADOOP}, \textsc{HDFS}, \textsc{YARN}, \textsc{HDDS} for our studies.
Although popular open-source projects such as Spark and Cassandra contain a large volume of issue reports, we excluded them due to the fragmented nature of their CI infrastructure and metadata. Unlike ASF projects, these projects often use cross platforms (e.g., GitHub Actions, CircleCI) and do not rely on a centralized issue tracker like JIRA. 
As a result, obtaining a consistent and complete dataset, including issue reports, CI outcomes, and source code patches, becomes significantly more difficult. 
To ensure consistent data collection and to reliably link CI build results with corresponding development activities, we limit our study to the seven ASF projects.
Although seven projects represent a limited number compared to many empirical studies, especially in the area of mining software repositories, we treat each project as a separate case study. For example, this number of projects enables us to delve deeper and manually investigate why developers consider the build failures to be unrelated. Additionally, it allows us to build models for each individual project instead of creating a general model based on a large amount of data. Our objective is to showcase the viability of our approach in identifying unrelated build failures, rather than striving for an extensive level of generalizability in our findings at this stage.

By applying the heuristics labeling to our seven studied projects, we obtain a set of \totalNumObj potentially unrelated build failures. 
We then use this list of potentially unrelated build failures to curate a representative sample of \ranSampNum~\textbf{truly unrelated build failures} with a 95\% confidence level and 5\% confidence interval (more on the dataset curation in Section~\ref{methodology}). By \textbf{truly unrelated build failures} we mean build failures that we manually confirmed to be deemed as unrelated by developers.

\begin{figure}[]
    \centering
    \includegraphics[width=0.9\textwidth, height=5cm]{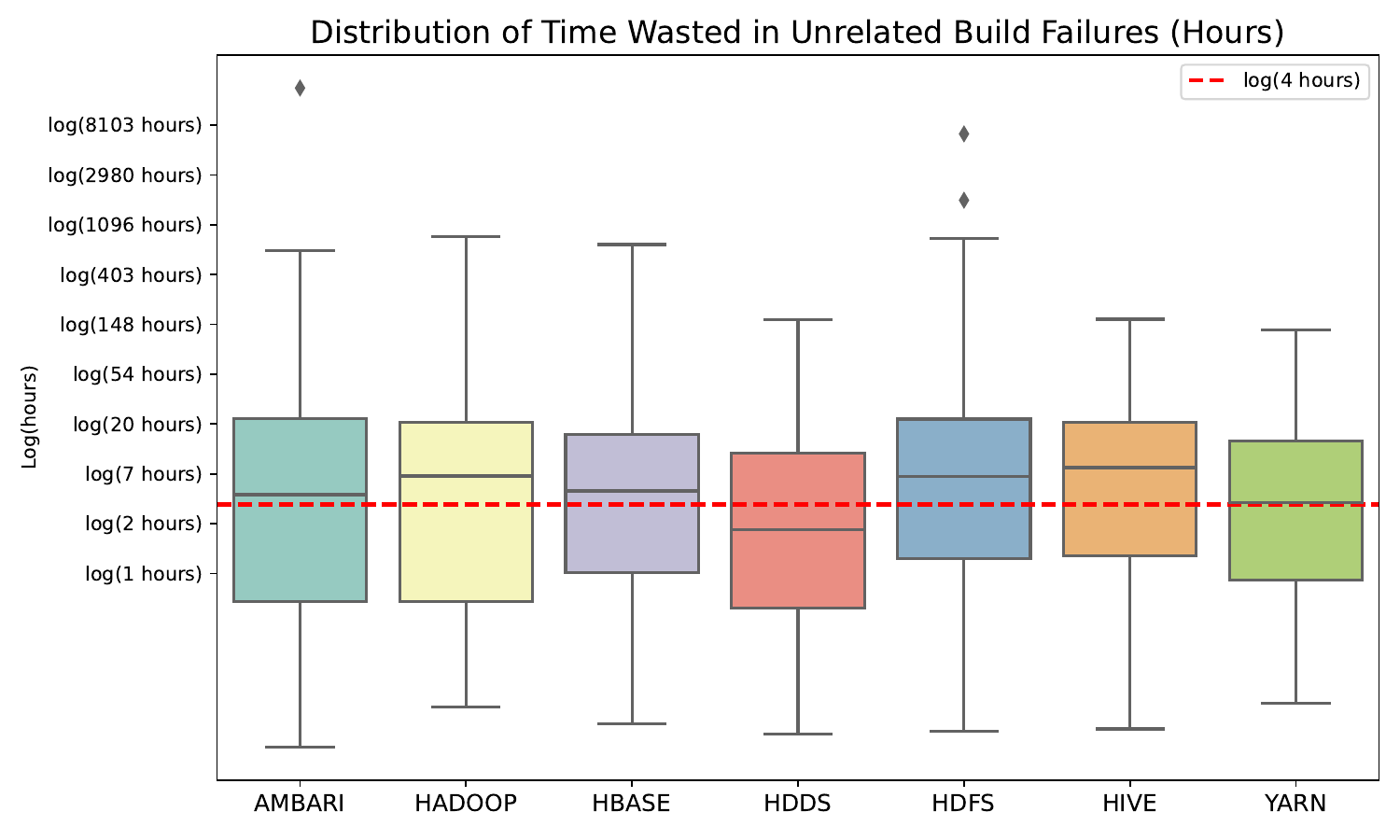}
    \caption{Time misspent on identifying unrelated build failures}
    \label{fig:results_time_wasted_ncrf}
\end{figure}

\begin{table*}[t]
\centering
\footnotesize
\begin{threeparttable}
\caption{Summary of Time Misspent and Code Complexity for unrelated build failures. Time Misspent measures the interval between CI failure notification and developer confirmation of ``unrelated''. Code Complexity metrics provide context for interpreting the time values.}
\label{tab:summary_ncrf}
\renewcommand{\arraystretch}{1.15}
\begin{tabular}{l r rrrr rr rr}
\toprule
& & \multicolumn{4}{c}{\textbf{Time Misspent (Hours)}} & \multicolumn{2}{c}{\textbf{Code Churn}} & \multicolumn{2}{c}{\textbf{Modified Files}} \\
\cmidrule(lr){3-6} \cmidrule(lr){7-8} \cmidrule(lr){9-10}
\textbf{Project} & \textbf{N} & \textbf{Median} & \textbf{Mean} & \textbf{Total} & \textbf{Std} & \textbf{Median} & \textbf{Mean} & \textbf{Median} & \textbf{Mean} \\
\midrule
AMBARI & 177   & 4.87  & 230.99 & 20,096 & 1,818 & 12   & 256  & 1 & 2.5 \\
HADOOP & 1,472 & 7.07  & 31.61  & 3,129  & 95    & 24   & 532  & 1 & 3.4 \\
HBASE  & 1,503 & 5.24  & 32.60  & 3,260  & 102   & 114  & 818  & 2 & 5.3 \\
HDDS   & 210   & 2.43  & 16.70  & 1,637  & 33    & 250  & 504  & 5 & 6.7 \\
HDFS   & 2,936 & 6.99  & 118.11 & 11,811 & 701   & 104  & 384  & 2 & 4.1 \\
HIVE   & 2,368 & 8.37  & 18.89  & 1,851  & 28    & 82   & 990  & 2 & 7.9 \\
YARN   & 1,650 & 4.18  & 11.51  & 1,139  & 20    & 234  & 826  & 3 & 6.7 \\
\midrule
\textbf{Total} & \textbf{10,316} & \textbf{5.24} & \textbf{63.03} & \textbf{42,923} & \textbf{709} & \textbf{104} & \textbf{678} & \textbf{2} & \textbf{5.5} \\
\bottomrule
\end{tabular}
\begin{tablenotes}
\footnotesize
\item \textbf{N}: Number of unrelated build failures identified by our heuristic approach.

\item \textbf{Code Churn}: Total lines changed (added + deleted + modified) across source code, configuration, and test files.

\item \textbf{Modified Files}: Number of source code and configuration files changed in the triggering push.
\end{tablenotes}
\end{threeparttable}
\end{table*}

To measure the time misspent on unrelated failures, we automatically scan all post-failure comments for a set of developers using the keywords (\eg~``not related'', ``unrelated'', ``irrelevant'', etc.,) which we derived from an initial sample of true unrelated failures. When one of these keywords appears in a comment, we take that moment as the explicit confirmation that the failure was deemed unrelated. For each confirmed unrelated failure, we record the interval between the timestamp of the CI bot’s initial failure notification comment and the first developer comment containing one of our unrelated keywords set.
We also measure the code churn of the push triggering the unrelated failure to understand the amount of code potentially withheld due to the unrelated failure.  \Fig \ref{fig:results_time_wasted_ncrf} shows that, on the median, the time required from developers to determine that a failure is unrelated is around 4 hours.
This observation highlights the importance of automatically identifying whether a build failure is unrelated, helping developers with making the decision whether they should proceed with the merge or not. Additionally, knowing whether a build is unrelated beforehand may also help developers debug the breakage more quickly since the problem scope to be investigated would be reduced (i.e., the breakage is unrelated to the current push).

To automatically identify unrelated build failures, we begin by investigating the ratio of potentially unrelated build failures to build failures in our larger dataset (i.e., not the manually curated dataset). We observe that the ratio of potentially unrelated build failures to build failures is $\frac{\totalNumObj}{\totalNumFails}\approx13.33\%$, meaning that the data is imbalanced. Additionally, it is also likely that there are several occasions where developers do not explicitly state when a build failure is unrelated. For instance, developers could have discussed the relatedness of a build failure on other communication channels~\cite{el2021exploring,ehsan2020empirical} as opposed to posting comments on the issue report. As such, it is likely that many positive cases (i.e., unrelated failures) cannot be retrieved from the available data. As such, both issues of i) imbalanced data and ii) unavailability of positive cases, motivate us to build a semi-supervised PU-learning approach, which is suitable for such challenges~\cite{liu2002partially,li2003learning,yu2002pebl,nguyen2011positive,yang2012positive,li2010positive}. We detail our research methodology in the following section.

%% file: src/tables/method_tbl1.tex
{\fontsize{8}{11}\selectfont
\begin{table}[htbp]
\centering
\caption{Project Metadata and CI/CD Failures. The selected projects are highlighted in red.}
\label{table:project_metadata_ci_cd}
\resizebox{0.8\linewidth}{!}{%
\begin{tabular}{m{.15\linewidth}m{.13\linewidth}m{0.1\linewidth}m{0.11\linewidth}m{0.15\linewidth}m{0.15\linewidth}}
\toprule
\textbf{Project Name} & \textbf{Num of Issues} & \textbf{Languages} & \textbf{Hist CI Builds} & \textbf{Ratio of Issues Contain Build Comments} & \textbf{Ratio of Heuristic Unrelated build Failures}
\\
\midrule
SPARK & 24463 & Scala & & & \\
\highlight{AMBARI} & \highlight{21936} & \highlight{Java} & \highlight{YES} & \highlight{32.35\% (\textgreater{}10\%)} & \highlight{9.1\%}\\
ARROW & 12514 & C++ & & &\\
CAMEL & 11652 & Java & & &\\
CASSANDRA & 10672 & Java & & &\\
BEAM & 7927 & Java & & &\\
IGNITE & 7404 & TypeScript & & &\\
\highlight{HBASE} & \highlight{7327} & \highlight{Java} & \highlight{YES} & \highlight{44.45\% (\textgreater{}10\%)} & \highlight{10.62\%}\\
NIFI & 6636 & Java & YES & 0.03\% &\\
KAFKA & 6505 & Java & & &\\
\highlight{HDFS} & \highlight{6421} & \highlight{JAVA} & \highlight{YES} & \highlight{53.22\% (\textgreater{}10\%)} & \highlight{17.52\%}\\
IMPALA & 6379 & C++ & YES & 0.04\% &\\
MESOS & 5310 & C++ & & &\\
KARAF & 4972 & Java & & &\\
\highlight{HDDS} & \highlight{4780} & \highlight{Java} & \highlight{YES} & \highlight{20.91\% (\textgreater{}10\%)} & \highlight{11.11\%}\\
AMQ & 4684 & Go & & &\\
\highlight{HADOOP} & \highlight{4562} & \highlight{Java} & \highlight{YES} & \highlight{46.62\% (\textgreater{}10\%)} & \highlight{17.70\%}\\
\highlight{YARN} & \highlight{4449} & \highlight{Java} & \highlight{YES} & \highlight{63.15\% (\textgreater{}10\%)} & \highlight{16.73\%}\\
WICKET & 4323 & Java & & &\\
\highlight{HIVE} & \highlight{4273} & \highlight{Java} & \highlight{YES} & \highlight{40.68\% (\textgreater{}10\%)} & \highlight{9.94\%}\\
\bottomrule
\end{tabular}}
\end{table}}

%% file: src/Methodology.tex
\textbf{Data Collection (DC).} Following the project selection, we begin to collect further data from JIRA, including information from issue reports (e.g., issue priority, issue activities), pushes linked to each issue report, and comments associated with issue reports (including developer and CI bot comments). As briefly explained in Section~\ref{motivating:example}, we develop a heuristic labeling approach to identify whether a CI build failure (as reported by the CI bot comments) is likely unrelated. First, for each issue report, we locate the CI bot comments based on the author of the comment (e.g., `Hadoop QA'). If the bot comment reports a build failure, our approach searches for relevant keywords in the following comments, such as ``\textbf{unrelated}'', or ``\textbf{irrelevant}'', that can indicate that the build failure is unrelated to the current push. We are aware that a heuristic-based approach such as this may likely yield sub-optimal accuracy. We accept this limitation at this stage, since our main goal is to strategically reduce our search space to then create a curated dataset, using the identified potentially unrelated failures as the initial input. Using the heuristic-based approach, we obtained \totalNumObj potentially unrelated build failures across our studied projects.

\input{src/tables/method_tbl2.tex}

\newcommand{\DependencyChainErrors}{\textbf{{concurrent issues}}}
\newcommand{\themeamount}{14 }

\textbf{Document Analysis (DA).} 
Once we obtain our set of potentially unrelated build failures in the DC step, we aim to further process the comments left by developers and understand the reasons why build failures are deemed as unrelated. The first step of this analysis is to build a statistically representative sample of {\bfseries truly} unrelated failures (as opposed to {\bfseries potentially} unrelated failures). For this purpose, we initially create a representative random sample of potentially unrelated failures across our seven projects. Considering a confidence level of 95\% and a confidence interval of 5\% on \totalNumObj unrelated build failures, we determine a sample size of \ranSampNum. To populate our sample, we initially verify whether the build failures identified as potentially unrelated by the HL approach are indeed unrelated. We do this by manually confirming that the HL-identified comment following a build failure indicates that the failure is unrelated (i.e., a truly unrelated build failure). Whenever we identify a false positive\footnote{The comment from developers included keywords suggesting doubt about the build failure being related to the code push. However, this was later confirmed by others to indeed be related to the code push.}, we replace it with another sample until we find one that is truly unrelated. 
To preserve the randomness of the process, whenever a false positive is detected, it is returned to the pool, and a new instance is randomly drawn from the entire pool of 10,316 instances. 
This verification and replacement procedure is repeated until we obtain 371 truly unrelated failures.
Two authors (\ie A.H. and M.M.)~worked together throughout the entirety of this step.

Once we assemble our sample, our goal is to further analyze the comments of developers to better understand why unrelated build failures are deemed as such. In general terms, comments from developers can be interpreted as digital documents describing a historical process behind an organization or project. As such, to elicit the knowledge embedded in these digital documents, we use a technique called document analysis~\cite{bowen2009document,zina2021essential}, a qualitative approach to examine documents and yield data from them. This data, which can be quotations, excerpts, or entire passages, are then organized into major themes along with examples. We treat the CI bot comments (i.e., the build failure logs), their related information, and following discussions as the digital documents to be inquired in our document analysis. The {\bfseries guiding question} for our document analysis is the following: {``\textit{Why do developers deem a build failure as unrelated to the current push}?''}

\begin{figure}[!h]
    \centering
    \includegraphics[width=0.9\textwidth]{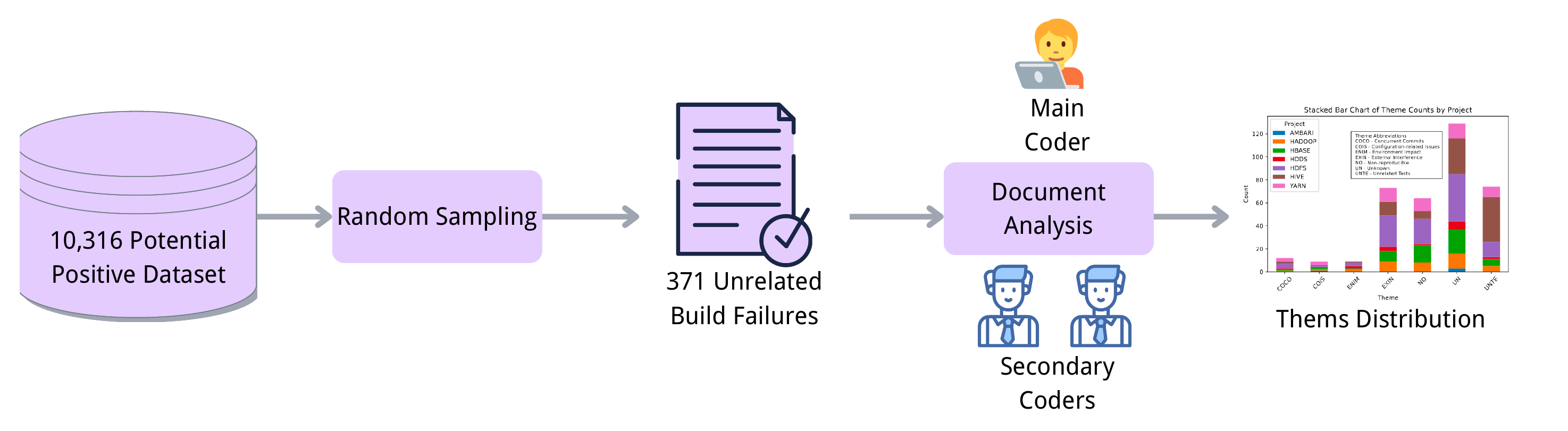}
    \caption{An overview of the document analysis process}
    \label{fig:results_rq1_ov}
\end{figure}

\Fig \ref{fig:results_rq1_ov} shows an overview of the document analysis process.
Three coders are involved in the document analysis process, all of whom have academic and industry backgrounds in software engineering. 
First, the first author (\ie ``main coder'' in \Fig \ref{fig:results_rq1_ov}) manually scrutinized each instance of our digital documents to answer the {\bfseries guiding question} for each of these instances. The answer to the {\bfseries guiding question} should be the reason why an unrelated build failure is deemed as such. For example, Eli Collins, a developer examining a build failure for issue HDFS-3582,\footnote{https://issues.apache.org/jira/browse/HDFS-3582} states: {``\textit{Fixed one findbugs warning (FindBugs isn't able to figure out one statement after terminate is currently unreachable). The other two are HDFS-3615. Test failures are unrelated.}''} The fact that the failed tests are associated with other pushes made for different issues, leads us to 
identify the \textbf{external interference} theme as the main reason contributing to the unrelated nature of the build failure. 
This inductive approach yields a set of themes that express the common reasons why developers deem build failures as unrelated to the present push. 
These themes will be presented and analyzed in Section~\ref{results}.

After generating and documenting the complete set of themes through multiple reflective iterations, the main coder engaged in discussions with two other authors (referred to as ``secondary coders'') to refine the themes and their interpretations.
The main coder generated 7 themes with detailed explanations as shown in Section~\ref{results},~\ie \textbf{Results}.
Then, The secondary coders performed a deductive analysis, using the seven themes established by the main coder. That is, the secondary coders independently labeled all the \ranSampNum~instances using the set of themes established in the previous step.
During this stage, the secondary coders had the opportunity to propose new themes. 
There are 2 additional themes created by one of the secondary coders, i.e., Time Out Issue, and Dependencies.  However, after carefully discussion among coders, those two themes can be categorized as \textbf{environmental impact} and \textbf{external interference} respectively after we made the scope of themes more clear.
We also use Cohen's Kappa to explore the level of agreement between the main coder and the secondary coders~\cite{kitchenham2013systematic, dybaa2008strength}. Cohen’s Kappa is a statistical measure used to assess the agreement between the two raters on two raters on categorical data while correcting for chance agreement. It is widely used in software engineering studies that involve qualitative coding or manual classification (e.g., defect categorization, thematic analysis).
We used the ``cohen\_kappa\_score'' method from the scikit-learn package to calculate Cohen’s Kappa.
We observed disagreements in 70 cases (18.92\%), with a Cohen’s Kappa score of 0.76 between the main coder~(A.H.) and the secondary coder~(M.M), indicating substantial agreement [4]. 
Similarly, A.H. and the secondary coder~(D.A) disagreed in 67 cases (18.11\%), achieving a Kappa score of 0.77, also reflecting substantial inter-rater reliability~\cite{ajibode2025towards}.

Through iterative consensus discussions, we refined our themes by merging similar proposals (\eg multiple timeout-related ideas into a unified \textbf{environmental impact} category and consolidating overly narrow labels (\eg subsuming ``Jenkins failures'' under Configuration Issues).
Finally, all coders engaged in online plenary sessions totaling around 65 hours, where they discussed the generated themes and the labeling of instances. By the end of these plenary sessions, the new themes were merged, and labeling conflicts were resolved, resulting in the final set of themes. This final set of themes is described in Section~\ref{results}. 

\textbf{Feature Engineering (FE).}
After studying the reasons why developers deem certain build failures as unrelated in the \textbf{DA step}, we use this knowledge to engineer features for our intended machine learning approach~\cite{yue2024unsupervised,yue2024variational}. 
From build failures, we extract \totalFeaNum features based on issue reports, preceding comments, and the triggering push.
The feature boxes below present the details of the extracted features, including their description, calculation formulas, and the rationale behind each metric.

Consequently, we extracted 19 features from the issue report dimension, 3 features from the comments dimension, and 11 features from the push dimension. Further details of the applied features are provided in the boxes below and in the Appendix.

\begin{figure}
   \centering
\includegraphics[width=0.8\textwidth]{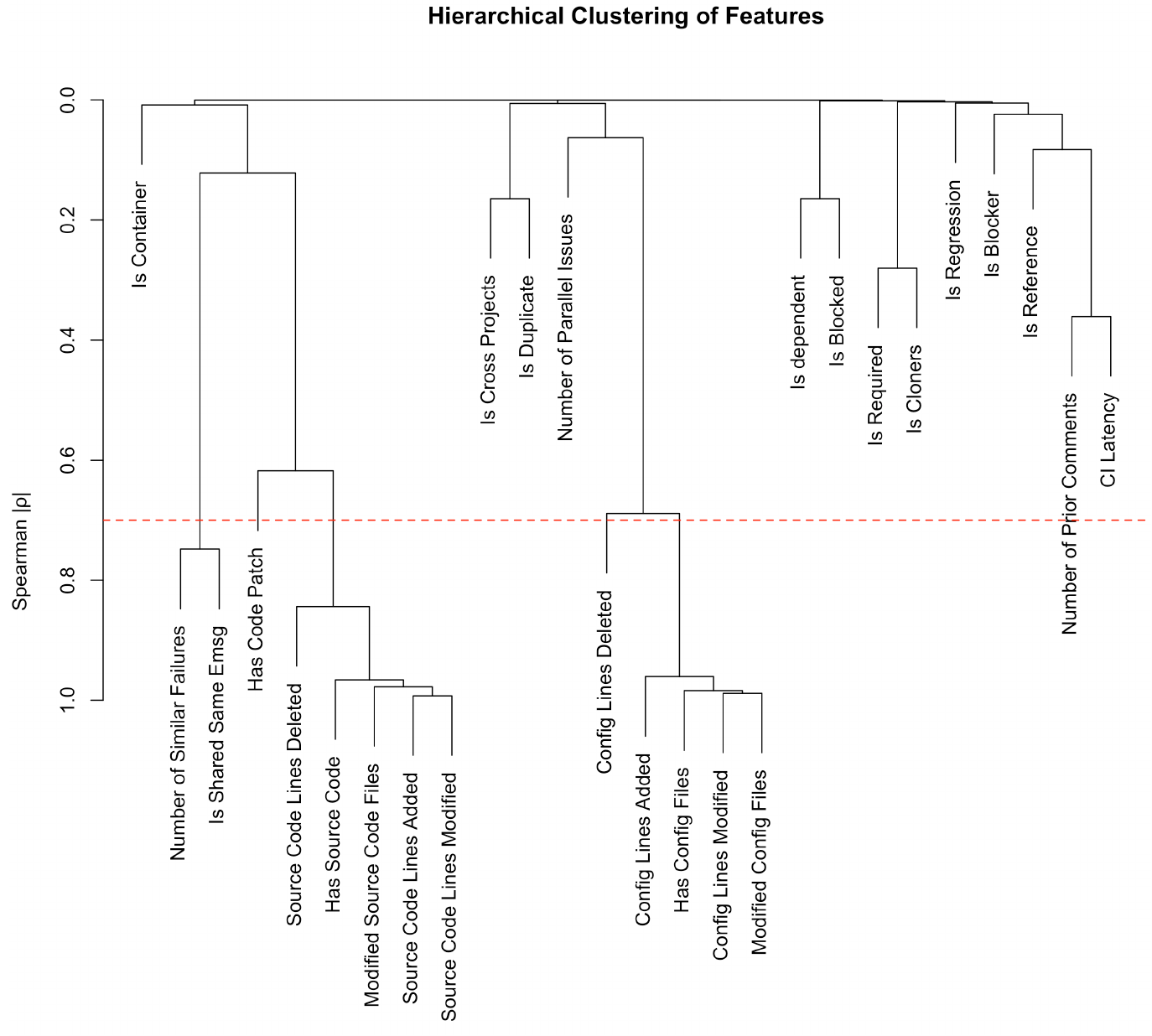}
   \caption{Dendrogram of hierarchical clustering based on Spearman correlation coefficients among features in the AMBARI project. The tree structure groups features by similarity, and the red dashed line marks the correlation threshold of $|\rho| = 0.7$. Features clustered above this threshold (below the red dashed line) were carefully reviewed, and less informative or redundant ones were removed to mitigate multicollinearity}
   \label{fig:spearman_corr}
\end{figure}

\input{src/tables/rq2_tbl1}
We also remove collinear features as they interfere with each other's significance within the model~\cite{huang2019empirical,mcintosh2018fix,mcintosh2016empirical}. We use Spearman’s \(\rho\) rank correlation on our \totalFeaNum features for each project.
Similar to previous work~\cite{huang2019empirical,ghaleb2019empirical,gousios2014exploratory}, we remove features if \(|\rho|\) $\ge 0.7$. 
In other words, if a pair of features exhibits a correlation \(|\rho|\) greater than 0.7, we prefer the simple and more informative metric over the complex metric~\cite{vandekerckhove201514,ghaleb2022studying}.
\Fig \ref{fig:spearman_corr} shows an example of Spearman’s rank analysis on features in AMBARI.
Based on the analysis, we retain the feature \textit{Number of Similar Failures} and remove \textit{Is Shared Same Emsg}, as the former provides a more informative signal.  
Similarly, we retain \textit{Modified Source Code Files} and discard highly correlated features including \textit{Source Code Lines Modified}, \textit{Source Code Lines Added}, \textit{Has Source Code}, and \textit{Source Code Lines Deleted}.
In the case of configuration-related features, we preserve \textit{Modified Config Files} while removing \textit{Config Lines Modified}, \textit{Has Config Files}, and \textit{Config Lines Added}, in accordance with the correlation results.
Further details on feature selection and Spearman’s rank correlation analysis for the remaining studied projects can be found in our replication package~\cite{githubrepo}. 

Also, if a feature can be predicted by other variables, it may lead to multicollinearity \cite{hanley1982meaning}. 
To mitigate multicollinearity, we conduct a redundancy analysis on the remaining features. We use the \(redun\) function in \(R\) to remove such redundant features if \(R^2\) \(>\) 0.9. Our \(redun\) analysis does not identify any redundant variables. We perform the correlation and redundancy filtering \emph{project-wise} because multicollinearity and redundancy are properties of the empirical distribution of a dataset rather than fixed properties of a feature.
Since our models are trained per project, each project can exhibit different feature variance and different sparsity of issue-link types, which may change the correlation structure among features.
Conducting Spearman’s rank correlation (with the $|\rho| \ge 0.7$ threshold) and the redundancy analysis per project, therefore, ensures that we remove only features that are redundant \emph{in that project}, avoiding biased removals that could be dominated by the largest project if all projects were aggregated.
Importantly, project-wise filtering does not imply completely different feature sets across projects. In our implementation, a substantial core set of features is consistently retained across all seven projects (\eg~\textit{CI Latency}, \textit{Number of Similar Failures}, and \textit{Number of Prior Comments}).

\textbf{Positive Unlabeled Learning (PUL).} Although a supervised binary classification approach might seem suitable, our dataset poses a key challenge: Positive instances (\ie unrelated build failures) cannot be reliably identified with sufficient completeness or accuracy. 
In terms of completeness, developers do not always confirm whether a build failure is unrelated to code changes, leaving many true positives hidden among the non-positive samples. 
In terms of accuracy, our heuristically labeled positives may include false positives, meaning that not all labeled unrelated failures are guaranteed to be correct. 
To illustrate, consider the issue of under-reporting smoking among pregnant populations~\cite{graham2003there}, where a significant number of smokers feel pressured to identify as non-smokers. In such cases, trusting the negative instances (i.e., non-smokers) is unreliable due to the presence of positive instances within them. The problem of unrelated build failures shares a similarity in that developers may not consistently and explicitly declare that a build failure is (un)related. Since the reliability of negative instances cannot be guaranteed, because the absence of evidence that a build failure is unrelated does not imply that the failure is truly related, yet we do have access to a trustworthy subset of positive instances, this setting is well suited for developing an approach based on Positive-Unlabeled (PU) learning~\cite{elkan2008learning,fusilier2015detecting,hsieh2015pu}.
We treat each of our studied projects as a case study and build a model for each of them.

In our scenario where the accuracy of deciding a build failure is related to the current code push (negative instances) is unreliable, however, a subset of unrelated build failures (positive instances) is available with high confidence, as 371 positive samples were manually investigated by coders in \textbf{DA} process.
Positive-unlabeled (PU) learning provides an effective framework for addressing the challenge. It is an approach that addresses scenarios where only a small subset of positive instances is labeled, and the rest of the data is unlabeled, potentially containing both positive and negative examples. 
To build our semi-supervised learning models, 
we adopt two distinct variants of the PU classifier, i) the Classic Elkan \& Noto classifier and ii) the Weighted Elkan \& Noto classifier~\cite{elkan2008learning}. 
The classic and weighted Elkan \& Noto classifiers operate under assumptions of \textit{SCAR}. 
The Selected Completely at Random (\textit{SCAR}) assumption~\cite{teisseyre2024verifying} posits that among all the positive examples in our dataset, each one has an equal and unbiased chance of being labeled, regardless of its specific features or attributes. 
In simpler terms, the labeling of positive examples happens entirely by chance and is not influenced by any characteristics of the data.
Applying this assumption allows us to treat the randomly chosen and labeled positive examples in \textbf{DA} as a representative sample of all positive instances.
It enables us to build a predictive model to learn from this unbiased sample and estimates the true likelihood of an example being positive, even when faced with a large number of unlabeled or ambiguous cases.
Formally, the SCAR can be expressed as 
\begin{equation}
p(s = 1 \mid x, y = 1) = p(s = 1 \mid y = 1) = c
\label{eq:scar}
\end{equation}
where:
\begin{itemize}
    \item \( s = 1 \) denotes a sample being labeled,
    \item \( y = 1 \) denotes a sample being truly positive,
    \item \( c = p(s = 1 \mid y = 1) \) is a constant probability, representing the chance that any positive example is labeled.
\end{itemize}
The classic Elkan \& Noto classifier assumes that unlabeled data can be treated as a general mixture of positive and negative instances.
To build the classic Elkan \& Noto model, it first trains a nontraditional classifier \(g(x)\) using the labeled positive examples and the unlabeled data.
This classifier estimates the probability that a sample is labeled given its features,  \(p(s = 1 \mid x)\).
The target is to learn a function \(f(x)\), where \(f(x) = p(y=1 \mid x)\).
Under the \textit{SCAR} assumption, the relationship between this probability of a sample is labeled \(p(s = 1 \mid x)\) and the true probability of an example being positive,  \(p(y = 1 \mid x)\), is given by:
\begin{equation}
f(x) = p(y = 1 \mid x) = \frac{p(s = 1 \mid x)}{c} = \frac{g(x)}{c}
\label{eq:classic_pu}
\end{equation}
By applying this method, the Classic Elkan \& Noto classifier effectively learns from the available data to predict the true probability  \(p(y = 1 \mid x)\), even in the absence of negative labels.
The Weighted Elkan \& Noto classifier extends the Classic Elkan \& Noto classifier to account for situations where the unlabeled data may not perfectly represent the negative class.
The key insight of this method is that the unlabeled data cannot be treated as entirely negative. 
Instead, each unlabeled example is viewed as both a potential positive and negative instance, with weights assigned to each interpretation.
The weights are derived using the probability of an example being positive given it is unlabeled, denoted as  \(p(y = 1 \mid x, s = 0) \).
These weights are computed based on the relationship between the probabilities of labeling and the underlying class distribution, expressed as:
\begin{equation}
w(x) = \frac{(1 - c)}{c} \cdot \frac{p(s = 1 \mid x)}{1 - p(s = 1 \mid x)}
\label{eq:classic_pu}
\end{equation}
Here,  \(c = p(s = 1 \mid y = 1)\)  represents the probability of labeling for a positive instance, a constant under the \textit{SCAR} assumption that positive examples are selected completely at random.
During the training process, each sample is assigned with two roles, i) positive example with weight \(w(x)\) and ii) a negative example with complementary weight \(1- w(x)\).
These two PU Learning methods play a crucial role in scenarios where collecting negative examples is difficult or infeasible. 
We can leverage only positive and unlabeled data, and develop effective predictive models for positive data (which are build failures unrelated to code pushes in this paper). 

\begin{figure}[]
   \includegraphics[width=0.9\textwidth]
   {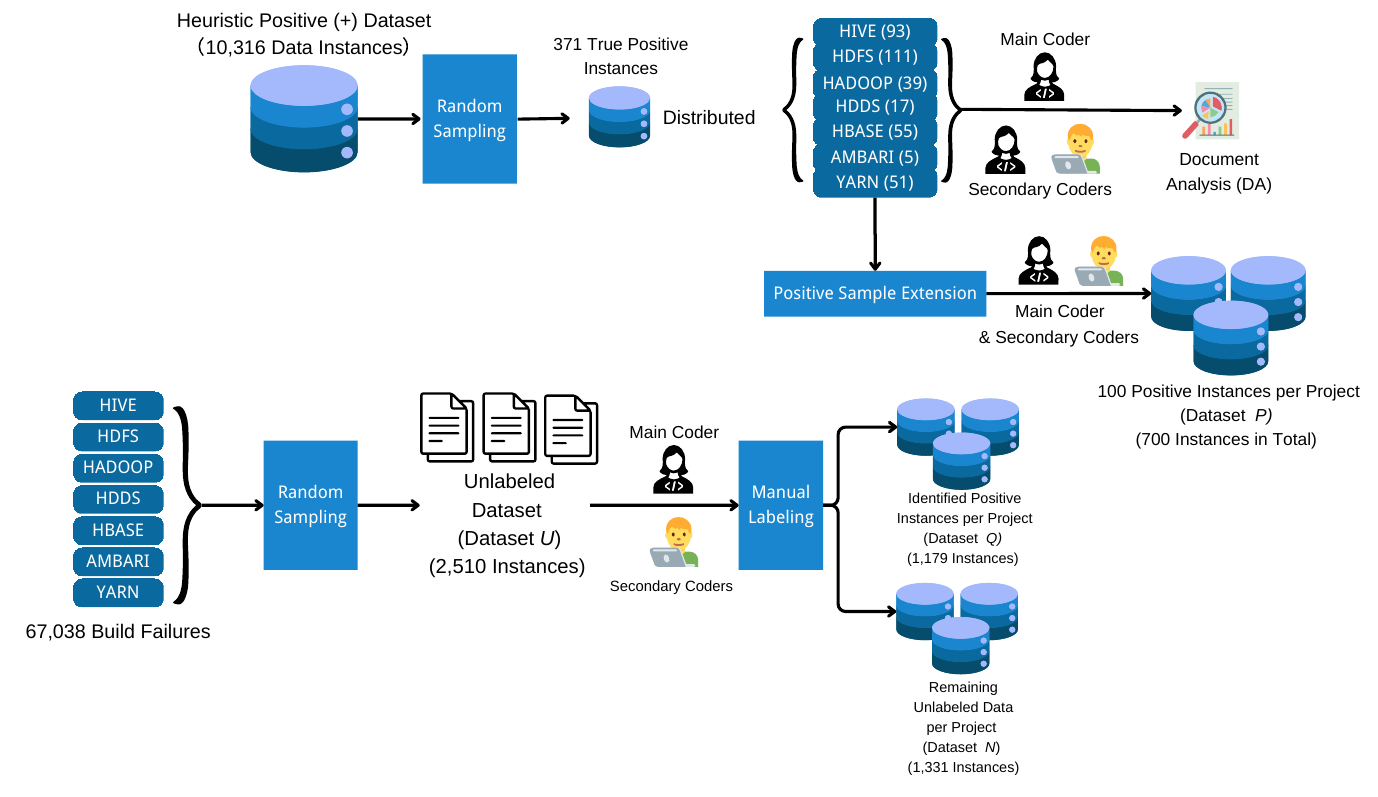}
   \caption{The Process Flow for Constructing the P, Q, and N Datasets.}
   \label{method:fig7}
\end{figure}

Evaluating models based on PU learning poses a challenge due to the absence of truly negative instances.
We only know the true class
label of a sample of positive instances (\ie \ranSampNum~truly unrelated build failures that are used in \textbf{DA}). 
The remaining build failures are not determined whether is related to code push.
Normally, the presence of unlabeled instances would not be possible to compute metrics such as true positive rate, false negative rate, precision, recall, and F-measure~\cite{japkowicz2011evaluating}. 
To address this, existing studies~\cite{japkowicz2011evaluating,elkan2008learning} commonly adopt an approach where a subset of positive instances is intentionally hidden within the negative set, thereby constructing an unlabeled set in which true negatives and hidden positives are treated as `unlabeled'.
Hence, the model is trained on PU data but evaluated on fully labeled data, and evaluation metrics, like precision and recall, can be accurately calculated.

\textbf{Progress for Constructing Evaluation Datasets P, Q, N. }~\Fig\ref{method:fig7} provides a visual representation of the dataset structure employed in this study, a standard procedure inspired by prior research in PU learning~\cite{elkan2008learning}. 
This methodology defines three distinct subsets of data: \(P\),  \(Q\), and \(N\). 
\(P\) is a subset consisting of explicitly labeled positive instances, representing data points that are definitively identified as positive based on prior knowledge~\cite{elkan2008learning}.
In the Heuristic-based labeling step, we obtain 10,316 positive samples, and we already have 371 samples from the \textbf{DA} step, which have already been confirmed as unrelated build failures (positive samples).
During the DA process, we established explicit, predefined criteria through dual-coder consensus. 
These criteria were consistently applied throughout the extended labeling process to ensure continuity and objectivity. 
The main coder in the DA step further manually labels each project to extend the positive sample size to 100 per project.
To ensure reliability, these 100 positive samples were independently validated by a second coder, and the two coders achieved a near-perfect Cohen’s Kappa of 0.98. The high agreement is due to the simplicity of the task, which involves checking whether developers explicitly stated that a failure is unrelated, leaving little room for interpretation.
As a result, we curated $100 * 7 = 700$ positive samples in total as the dataset \(P\). Those 700 positive samples overlap with the 371 samples used in the DA step. 
\begin{table}[htbp]
\centering
\begin{threeparttable}
\caption{Distribution of U, P, Q, and N among the seven studied projects}
\label{tab:distribution-u-p-q-n}
\renewcommand{\arraystretch}{1.15}

{\scriptsize
\setlength{\tabcolsep}{3pt}

\begin{tabular}{
  l
  S[table-format=5.0]
  S[table-format=4.0]
  S[table-format=1.3]
  S[table-format=3.0]
  S[table-format=3.0]
  S[table-format=3.0]
  S[table-format=3.0]
  S[table-format=3.0]
  S[table-format=1.3]
}
\toprule
& \multicolumn{3}{c}{\textbf{Heuristic labeling}} 
& \multicolumn{6}{c}{\textbf{Manul labeling}} \\
\cmidrule(lr){2-4} \cmidrule(lr){5-10}
\rowcolor{gray!10}
\textbf{Project} &
\multicolumn{1}{c}{\textbf{\begin{tabular}[c]{@{}c@{}}Total\\Samples\end{tabular}}} &
\multicolumn{1}{c}{\textbf{\begin{tabular}[c]{@{}c@{}}Heuristic\\Positive\\Sample\end{tabular}}} &
\multicolumn{1}{c}{\textbf{\begin{tabular}[c]{@{}c@{}}Unlabeled\\Ratio$^\dagger$\end{tabular}}} &
\multicolumn{1}{c}{\textbf{\begin{tabular}[c]{@{}c@{}}Sample\\Distribution\\in DA Step\end{tabular}}} &
\multicolumn{1}{c}{\textbf{U$^*$}} &
\multicolumn{1}{c}{\textbf{P}} &
\multicolumn{1}{c}{\textbf{Q}} &
\multicolumn{1}{c}{\textbf{N}} &
\multicolumn{1}{c}{\textbf{\begin{tabular}[c]{@{}c@{}}Unlabeled\\Ratio$^\ddagger$\end{tabular}}}
\\
\midrule
AMBARI & 1950  &  177 & 0.909 &   5 & 322 & 100 & 246 &  76 & 0.236 \\
HADOOP & 8649  & 1472 & 0.830 &  39 & 368 & 100 & 130 & 238 & 0.647 \\
HBASE  & 14160 & 1503 & 0.894 &  55 & 375 & 100 & 191 & 184 & 0.491 \\
HDDS   & 1878  &  210 & 0.888 &  17 & 320 & 100 & 149 & 171 & 0.534 \\
HDFS   & 16896 & 2936 & 0.826 & 111 & 376 & 100 & 166 & 210 & 0.559 \\
HIVE   & 23885 & 2368 & 0.901 &  93 & 379 & 100 & 127 & 252 & 0.665 \\
YARN   & 9936  & 1650 & 0.834 &  51 & 370 & 100 & 170 & 200 & 0.541 \\
\midrule
\rowcolor{gray!10}
\textbf{Total} 
  & \textbf{77354}
  & \textbf{10316}
  & \textbf{0.867}
  & \textbf{371}
  & \textbf{2510}
  & \textbf{700}
  & \textbf{1179}
  & \textbf{1331}
  & \textbf{0.530} \\
\bottomrule
\end{tabular}
} 

\vspace{0.3em}
\begin{tablenotes}[flushleft]\footnotesize
\item[$^*$] Dataset \(U\) is randomly selected from the total samples in each project at a 95\%\ confidence level with a 5\%\ confidence interval.
\item[$^\dagger$] The unlabeled ratio next to the heuristic positive samples is computed as 
\((\text{Total Samples} - \text{Heuristic Positive Sample}) / \text{Total Samples}\).
\item[$^\ddagger$] The unlabeled ratio in the rightmost column is computed as \((U - Q) / U\).
\end{tablenotes}
\end{threeparttable}
\end{table}

Dataset \(U\) consists of subsets \(Q\) and \(N\). We randomly select samples from each project’s full dataset for manual labeling. The sample size for each project was determined using a confidence level of 95\% with a confidence interval of 5\%, ensuring that the selected samples are representative of the overall dataset \(U\) for each project. 
During the manual labeling process, two coders independently reviewed the failure logs and follow-up developer comments to determine whether each failure was truly unrelated, achieving a near-perfect agreement with a Cohen’s Kappa of 0.91.
The labeled positive samples from \(U\) are curated as dataset \(Q\), and the rest of the data are called dataset N. Table~\ref{tab:distribution-u-p-q-n} shows the detailed number of randomly selected samples for each project. As a result, we obtained 1,179 samples in \(Q\) and 1,331 in \(N\), respectively.

\textbf{SCAR Assumption Validation.} Previous work [2] proposed a statistical test to verify whether a dataset satisfies the SCAR (Selected Completely At Random) assumption, which is essential for valid PU-learning. 
We adopt their method and apply it to the datasets of seven projects. 
In our setting, we treat the 700 positively labeled samples (identified by heuristics and manually confirmed) as known positives \(P\), while the remaining samples are considered unlabeled \(U\).
Following [2], we first approximate the class prior \(\pi\) (the proportion of positives in the entire dataset) using the ratio of manually confirmed positives Q to the unlabeled dataset U. A simple probabilistic classifier (e.g., RandomForest) is then trained with P as positives and a random subset of \(U\) (assumed mostly negative) as negatives, and applied to the rest of U to estimate the probability of being positive for each instance. Based on these scores and \(\pi\), we derive an estimated positive set. To mimic the SCAR mechanism, we randomly relabel instances within this estimated positive set with probability \(c=P(S=1)\), where \(P(S=1)\) is the empirical proportion of labeled positives among all estimated positives. 
This simulates a random selection of labeled positives under the SCAR assumption. 
We then compute a test statistic that measures the discrepancy between the known positives (\(P\)) and the estimated positives. 
Specifically, another classifier is trained to distinguish between the two groups, and the AUC is used as the statistic. 
A value close to 0.5 indicates similarity. 
To obtain a null distribution under the SCAR hypothesis, we repeat this procedure B=1000 times, and compare the observed statistic against this distribution to calculate a p-value. 
If the \(p-value\) exceeds a significance threshold (\eg 0.05), we fail to reject the null hypothesis and regard the SCAR assumption as plausible. 
Otherwise, we reject it. 
Across all seven projects, the resulting \(p-values\) are consistently greater than 0.05. This indicates that we do not find statistical evidence against the SCAR assumption in our datasets, suggesting that SCAR is a plausible working assumption and providing justification for applying PU learning in our study.

To evaluate the effectiveness of our PU-learning models, we compare them against three simple yet informative baselines: i) Random Model: This baseline randomly assigns labels (unrelated or related) to instances, serving as a reference point that reflects performance expected by chance. ii) This baseline naively predicts all instances as unrelated (\ie positive class). While simplistic, it serves as a useful comparison to assess how much benefit our model provides over always assuming unrelated failures. iii) Heuristic from Prior Error Messages (HPEM) approach, it aims to identify unrelated build failures by checking whether the error message matches that of the most recent failed build.

By following the evaluation approach inspired by evaluation settings in the prior work~\cite{elkan2008learning}, our experiments involve the comparison of our PU-learning models against 3 baselines: (i) Random Model on \(P\) versus \(Q \cup N\), (ii) Consistently Positive Model on \(P\) versus \(Q \cup N\), (iii) HPEM on \(P\) versus \(Q \cup N\), (iv) Classic Elkan \& Noto Classifier on \(P\) versus \(Q \cup N\), (v) Weighted Elkan \& Noto Classifier on \(P\) versus \(Q \cup N\). 
For each of the four methods, we perform 10-fold cross validation, and obtain a combined confusion matrix from the ten testing subsets.
With one confusion matrix for each approach, we compute its recall and precision, f1 score and AUC as evaluation metrics.
The rationale for using the confusion matrix is based on prior work~\cite{elkan2008learning}, which employed a combined confution matrix in their PU learning evaluation. Since the test set in each fold is relatively small, the resulting confusion matrices may be affected by sampling variability. To mitigate this issue, a combined confusion matrix is more suitable for evaluation. Particularly in scenarios where only positive labels are available, the aggregation of metrics provides a more stable and reliable basis for calculating evaluation and performance metrics~\cite{forman2010apples}.
Cross-validation proceeds as follows. 
Partition \textit{P}, \textit{Q}, and \textit{N} randomly into ten subsets each of size as equal as possible.
For each of the ten trials, reserve one subset of \textit{P}, \textit{Q},
and \textit{N} for testing.
In each trial number \( i \), this classifier is applied to the testing subsets \( P_i \), \( Q_i \), and \( N_i \), yielding a confusion matrix of the following form:
\[
\renewcommand{\arraystretch}{1.0} 
\begin{array}{c|ccc}
& \text{positive} & \text{negative} & \text{predicted} \\
\midrule
P_i \cup Q_i & a_i & b_i & \\
N_i & c_i & d_i & \\
\text{actual} & & &
\end{array}
\]
The combined confusion matrix reported for each approach has entries \( a = \sum_{i=1}^{10} a_i \) etc. In this matrix \( a \) is the number of true positives, \( b \) is the number of false negatives, \( c \) is the number of false positives, and \( d \) is the number of true negatives. Finally, precision is defined as \( p = \frac{a}{a + c} \) and recall as \( r = \frac{a}{a + b} \). The F-score is calculated as \(F_1 = 2 \cdot \frac{\text{p} \cdot \text{r}}{\text{p} + \text{r}}\). Each of the four models yields a classifier that assigns numerical scores to the test data points, so for each model, we can measure its AUC values by true positive rate (TPR, or sensitivity) against false positive rate (FPR, or 1 - specificity (or TNR)).

\textbf{Feature Importance (FI).}
After evaluating the performance of our models in the \textbf{PUL} step, we study which features play an important role in detecting whether a build failure is unrelated.  
To compute feature importance, we apply the \textit{permutation feature importance} method which is particularly beneficial when dealing with non-linear or complex estimators (\ie tree model) that lack transparency~\cite{Leo2001Random}.
Our based estimator in the classic and weighted PU classifier is Random Forest, which allows us to calculate each feature's permutation feature importance value based on the trained PU classifiers.
The fundamental idea behind permutation feature importance is to assess the contribution of a single feature by measuring how much the model's performance decreases when the values of that feature are randomly shuffled. In our case, we use recall and F1-score as the performance metrics. This technique has been widely used in previous studies~\cite{rajbahadur2017impact, tantithamthavorn2018impact, jiarpakdee2020impact, breiman2001random}. 

To compute the feature importance for each project, we collect feature importance values for each feature during each round of a 10-fold cross-validation process. For every fold, the corresponding feature importance scores are recorded. Once the cross-validation is finished, we have a set of feature importance values for each feature. We then calculate the median value of the feature importance values across these 10 rounds. By using the median, our ranking is less influenced by outliers and better reflects the central tendency of feature importance across all folds. Finally, we rank the features based on their median feature importance values, ordering them from the highest to the lowest to identify the most influential features in our model.

The computation of permutation importance allows us to identify which features contribute most significantly to the generalization capability of the examined model, not to mention that features that demonstrate importance on the training set but fail to do so on the held-out set could potentially lead to model overfitting. 

By examining feature importance, we gain insights into how each feature contributes to the model's decision-making process. Understanding the relative significance of different features helps us better understand the relationship between these features and unrelated build failures, providing potential insights for practitioners and researchers.

\textbf{Cross-Project Validation (CPV).} A practice of training a model on data from one or more software projects and then validating or testing it on entirely different projects~\cite{herbold2018comparative}. The goal of CPV is to assess how well our model can generalize across projects, which is especially helpful for transferring our PU Learning model to a new project that lacks sufficient historical data.
In our study, we adopt a leave-one-project-out design.
For each of the seven studied projects, we designate it as the testing project and use the data from the other remaining six projects for training. 
This protocol is widely used in prior research~\cite{zimmermann2009cross,turhan2009relative}, as it provides a strict generalizability test compared to within-project validation, ensuring that project-specific metrics do not inflate the performance of models.

%% file: src/tables/method_tbl2.tex

%% file: src/tables/rq2_tbl1.tex
\begin{featureblock}{Issue Report Level}{}{fig:issue_features}

\textbf{1. Priority} indicates the urgency or importance of an issue report, often assigned in Jira. \\
\textbf{Rationale:} Urgent issues may attract faster responses and multiple developers, increasing chances of overlapping or rushed changes that could lead to unrelated build failures.

\textbf{2.Number of Parallel Issues} quantifies the count of issues that are being worked on simultaneously within a project \cite{perry2001parallel}. 
  \par
  \par
  \begin{center}
      $\sum_{n=1}^{m} \left( \text{Count}(T_n = T_i) \right)$
  \end{center}
  $Where$:
    
    \begin{itemize}
    \item $m$ denotes the count of issue reports that were opened prior to the date when the current issue report $i$ was submitted.
    \item $T_n$ is the date of the $n_{th}$ issue report opened.
    \item $T_i$ is the date of the current issue report $i$ opened.
    \end{itemize}

\textbf{Rationale:} The concurrent work on issues may cause unrelated build failures as developers may not always communicate about the parallel changes being made.

\par
\textbf{3. Is Cross Projects} indicates whether an issue report is related to multiple projects within the organization.

\textbf{Rationale:} Changes touching multiple projects increase complexity due to dependencies, coordination needs, and integration contexts. This raises the likelihood of forgetting updates, resource conflicts, mismatched environments\cite{jayasuriya2025extended}.
\par

\textbf{4. Is Duplicate}, whether the issue is a duplicate of another issue
\par
\textbf{5. Is Blocker} whether the issue is preventing the resolution of another issue\par
\textbf{6. Is Blocked} whether the issue is being blocked or held up by another issue\par 
\textbf{7. Is Regression} whether the issue has already been resolved in an earlier version but has reappeared in the current version
\par
\textbf{8. Is Dependent} whether the issue is dependent on the resolution of another issue
\par
\textbf{9. Is Incorporates} whether the linked issue is incorporated or relied upon in some way by the current issue.
\par
\textbf{10. Is Required} whether the issue is dependent on the resolution of another issue.
\par
\textbf{11. Is Reference} whether the issue is simply mentioning or referring to another issue in some way.
\par
\textbf{12. Is Completes} whether the issue is linked to another issue that it completes or resolves.
\par
\textbf{13. Is Testing} whether the issue is linked to another issue to a testing or quality assurance (QA) task related to that issue.
\par
\textbf{14. Is Issue Split} whether the issue has been divided into two or more separate issues.\par
\textbf{15. Is Supercedes} whether the issue is linked to another issue that it replaces or supersedes. \par
\textbf{16. Is Cloner} whether the issue has been cloned in Jira
\par
\textbf{17. Is Container} indicates the relationship between issues and their respective location or grouping within the project.
\par
\textbf{18. Is Parent Feature} whether an issue represents a high-level concept that can be decomposed into smaller issues.\par
\textbf{19. Is Child-Issue } refers to the relationship between a parent issue and a child issue, known as a ``sub-task" relationship
\par
\textbf{Rationale:} Jira link types such as above features (4-19) establish interdependencies among issues. The improperly managed or ambiguous linkages can lead to unintended side effects, such as incomplete implementations, overlooked dependencies, or conflicts between changes, which can cause build failures in unrelated components\cite{luders2022beyond, luders2022automated,luders2019openreq}. 
\end{featureblock}

\begin{featureblock}{Historical Comment Level}{}{fig:comment_features}

\textbf{20. Number of Prior Comments} represents the number of prior comments associated with a build failure comment. 
\par
\begin{center} $\sum_{i < \text{curr}} \text{count}(C_i)$\end{center}
    \par
 $Where$:
\begin{itemize}
 \item $i$ iterates through the comments before the current failed CI. $curr$ is the index of the current build failure comment.
 \item $C_i$ is the comment at index $i$ in the historical comments.
 \item $count(C_i)$ is the count of comments by developers for each comment $C_i$
\end{itemize} 
\par
\textbf{Rationale:} Extensively discussed issues help evaluate build failure relevance to code changes.

\textbf{21. Is Shared Same Emsg} a boolean value that indicates whether a pair of build failures shares the same error message. 
\par
\begin{center}
     $ 
    \begin{cases}
       \text{true} & \text{if } C_{\text{curr}} \text{ is in } C_{\text{prev}} \\
        \text{false} & \text{otherwise}
    \end{cases}
    $
\end{center}
\par
$Where$:
\begin{itemize}
  \item \(C_{\text{curr}}\) is the error message from the current build failure.
  \item \(C_{\text{prev}}\) is the set of error messages from previous build failure comments.
\end{itemize}
\par
\textbf{Rationale:} A build failure that shares the same error message with previous build failures' comments is possibly not related to the current push as it can represent an error inherited by another breaking change. 

\textbf{22. Number of Similar Failures} represents the count of build failures that have an intersection with the set of failed test/exception classes extracted from the current build's failure messages.
\par
  \begin{center}
      $\sum_{i=1}^{n} |F_{curr} \cap F_{pre_{i}}|$
  \end{center}
  \par
$Where$:
\begin{itemize}
  \item $F_{pre_{i}}$ is the set of failed test/exception class cases from the failure messages of the $i_{th}$ previous build, where $i$ varies from $1$ to $n$, $n$ is the number of previous builds.
  \item $F_{curr}$ is the set of failed test/exception class cases from the failure messages of the current build.
  \item $n$ is the number of previous builds.
\end{itemize} 
\par
\textbf{Rationale:} If a build failure shares similar failed classes with previous failures, it may be caused by breaking changes unrelated to the current push. 
    
\end{featureblock}

\begin{featureblock}{Push Level}{}{fig:push_features}
\textbf{23. CI Latency} refers to the duration between when a code patch is uploaded/pushed and when the build is triggered for that code patch.\par
\textbf{Rationale:} Long CI latency may cause unrelated failures, as it becomes harder to associate code changes with specific build outcomes.

\textbf{24. Has Code Patch} a boolean flag that indicates whether a build failure includes the corresponding code patch with the failed execution. If the CI build comment includes a code patch file, the value is $true$ otherwise, it is $false$.
\par
\textbf{Rationale}: Without a code patch, it becomes more challenging to pinpoint the code changes responsible for the build failures.

\textbf{25. Has Config Files}
A boolean flag indicating whether a code patch contains modifications in files ending with \texttt{.yaml}, \texttt{.xml}, or \texttt{.properties}~\cite{ghaleb2019empirical}.
\par
\textbf{Rationale}: Changes in configuration files are more likely to introduce build failures due to misconfiguration. If no such files are modified, we can reasonably exclude configuration-related causes of failure.

\medskip

\textbf{26. Config Lines Deleted}
The number of lines deleted from configuration files (\texttt{.yaml}, \texttt{.xml}, \texttt{.properties}) in a code patch~\cite{ghaleb2019empirical}.
\par
\textbf{Rationale}: Deletions in config files can remove critical settings or dependencies, potentially causing unexpected failures.

\medskip

\textbf{27. Config Lines Added}
The number of lines added to configuration files in a code patch~\cite{ghaleb2019empirical}.
\par
\textbf{Rationale}: Additions in config files may include new dependencies or settings that unintentionally break other parts of the system.

\medskip

\textbf{28. Config Lines Modified}
The number of lines modified in configuration files in a code patch~\cite{ghaleb2019empirical}.
\par
\textbf{Rationale}: Modifications in config files can subtly change behavior or dependencies, possibly triggering unrelated build failures.

\medskip

\textbf{29. Has Source Code}
A boolean flag indicating whether a code patch contains modifications in files ending with \texttt{.java}~\cite{ghaleb2019empirical}.
\par
\textbf{Rationale}: The absence of source code changes strengthens the suspicion that failures originate from configuration rather than code logic.

\medskip

\textbf{30. Source Code Lines Added}
The number of lines added to Java source files in a code patch~\cite{ghaleb2019empirical}.
\par
\textbf{Rationale}: Large additions in source code can introduce incomplete or experimental features, leading to failures not directly related to other ongoing work.

\medskip

\textbf{31. Source Code Lines Deleted}
The number of lines deleted from Java source files in a code patch~\cite{ghaleb2019empirical}.
\par
\textbf{Rationale}: Removing code may eliminate functionality or dependencies, which can unexpectedly affect unrelated builds.

\medskip

\textbf{32. Source Code Lines Modified}
The number of lines modified in Java source files in a code patch~\cite{pan2021continuous,saidani2020predicting,ghaleb2019empirical}.
\par
\textbf{Rationale}: Intense modifications across many lines can obscure the cause-and-effect relationship, making it harder to rule out unrelated code as the failure reason.

\medskip

\textbf{33. Modified Source Code Files}
The count of Java files modified in a code patch.
\par
\textbf{Rationale}: When only a few files are modified, it's easier to identify the source of failure; many files modified complicate analysis and increase the risk of unrelated build failures.
    
\end{featureblock}

%% file: src/results.tex
\newcommand{\ConcurrentCommits}{\textbf{{concurrent commits}}}
\newcommand{\ConfigurationrelatedIssues}{\textbf{{configuration-related issues }}}
\newcommand{\CrossScenario}{\textbf{{Cross Scenario }}}
\newcommand{\ExternalInterference}{\textbf{external interference }}
\newcommand{\ExternalResourceDependency}{\textbf{{External Resource Dependency }}}
\newcommand{\ExternalEnvironmentImpact}{\textbf{{External Environment Impact }}}
\newcommand{\FunctionDescriptionChange }{\textbf{{Function Description Change }}}
\newcommand{\LogrelatedErrors }{\textbf{{Log-related Errors }}}
\newcommand{\Nonreproducible}{\textbf{{non-reproducible}}}
\newcommand{\RebaseRequired }{\textbf{{Rebase Required }}}
\newcommand{\ResilientTestCode }{\textbf{{Resilient Test Code }}}
\newcommand{\UnrelatedTestingClasses}{\textbf{{unrelated tests}}}
\newcommand{\VersionSpecificErrors }{\textbf{{Version-Specific Errors }}}
\newcommand{\Unknown }{\textbf{{unspecified by the developer }}}

\subsection{Developer reasons for deeming CI failures as unrelated.}

\input{src/tables/rq1_tbl2}

\begin{figure}[!h]
    \centering
    \includegraphics[width=0.8\textwidth]{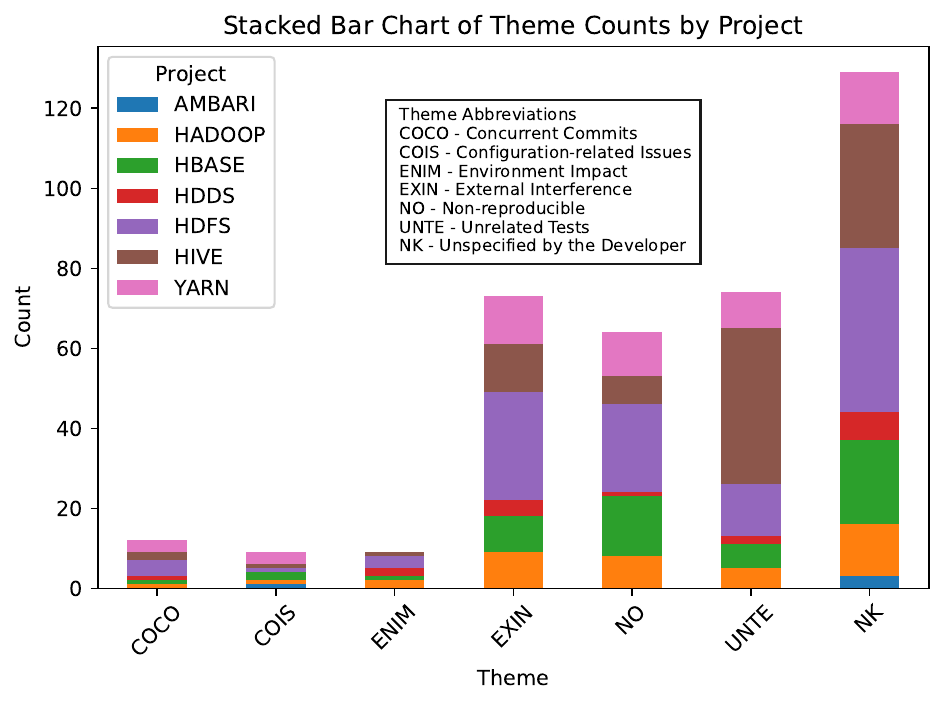}
    \caption{The distribution of representative samples (371) across each theme and project. The x-axis represents the abbreviated name of the themes, and the y-axis shows the number of samples in each theme.}
    \label{fig:rq1_theme_dist}
\end{figure}

\input{src/tables/rq1_tbl3}
\Tbl~\ref{results:rq1_tbl2} shows the themes generated in our thematic analysis during the DA step along with their descriptions.  \Fig \ref{fig:rq1_theme_dist} shows the distribution of themes across the studied projects. 
Unfortunately, a significant portion (34.86\%, $\frac{129}{371}$) of unrelated failures are deemed as such for reasons {\bfseries \Unknown } to us, i.e., developers simply state that {``\textit{the failure is unrelated}''} without providing further context or details. 

On the other hand, the \UnrelatedTestingClasses~ (20\%, $\frac{74}{371}$) theme is the second most-frequent reason for developers deeming a failure as unrelated. This is the case when developers check that the failing tests are unrelated to the scope of the current push. For instance, in issue YARN-1410,\footnote{\url{https://issues.apache.org/jira/browse/YARN-1410}} a developer declares that the {``\textit{test case failure is not related}.''} Such cases may occur due to several reasons, the presence of previous and unfixed broken builds, which would go against CI best practices but still occur in practice~\cite{duvall2007continuous,felidre2019continuous}.
The third and fourth most frequent reasons for deeming build failures as unrelated are represented by the \ExternalInterference~ (19\%, $\frac{73}{371}$) and \Nonreproducible~ (17\%, $\frac{64}{371}$) themes, respectively.
The \ExternalInterference~ theme describes build failures influenced by factors beyond the current code changes. These may include code pushes intended to address issues from other tasks or external modifications to dependencies on external resources.
The \Nonreproducible~ theme represents situations when developers, to the best of their efforts, cannot reproduce the failure that is reported by the CI bot. These failures may occur due to external factors, such as differences in environments, system load, and more.

As \Fig\ref{fig:rq1_theme_dist} shows, the other themes are not as frequent as the top four themes mentioned above. However, they also represent interesting scenarios, such as the \ConfigurationrelatedIssues~theme, which represents situations where the failure is associated to faulty configuration settings that are unrelated to the current push.

Overall, our themes highlight that the challenges associated with unrelated build failures are tightly related to the concurrent nature of software development where several changes to the software are applied concurrently.

\subsection{PU-Learning Approach to Detect Unrelated Build Failures}

\begin{figure}[h]
    \centering
    \includegraphics[width=\textwidth]{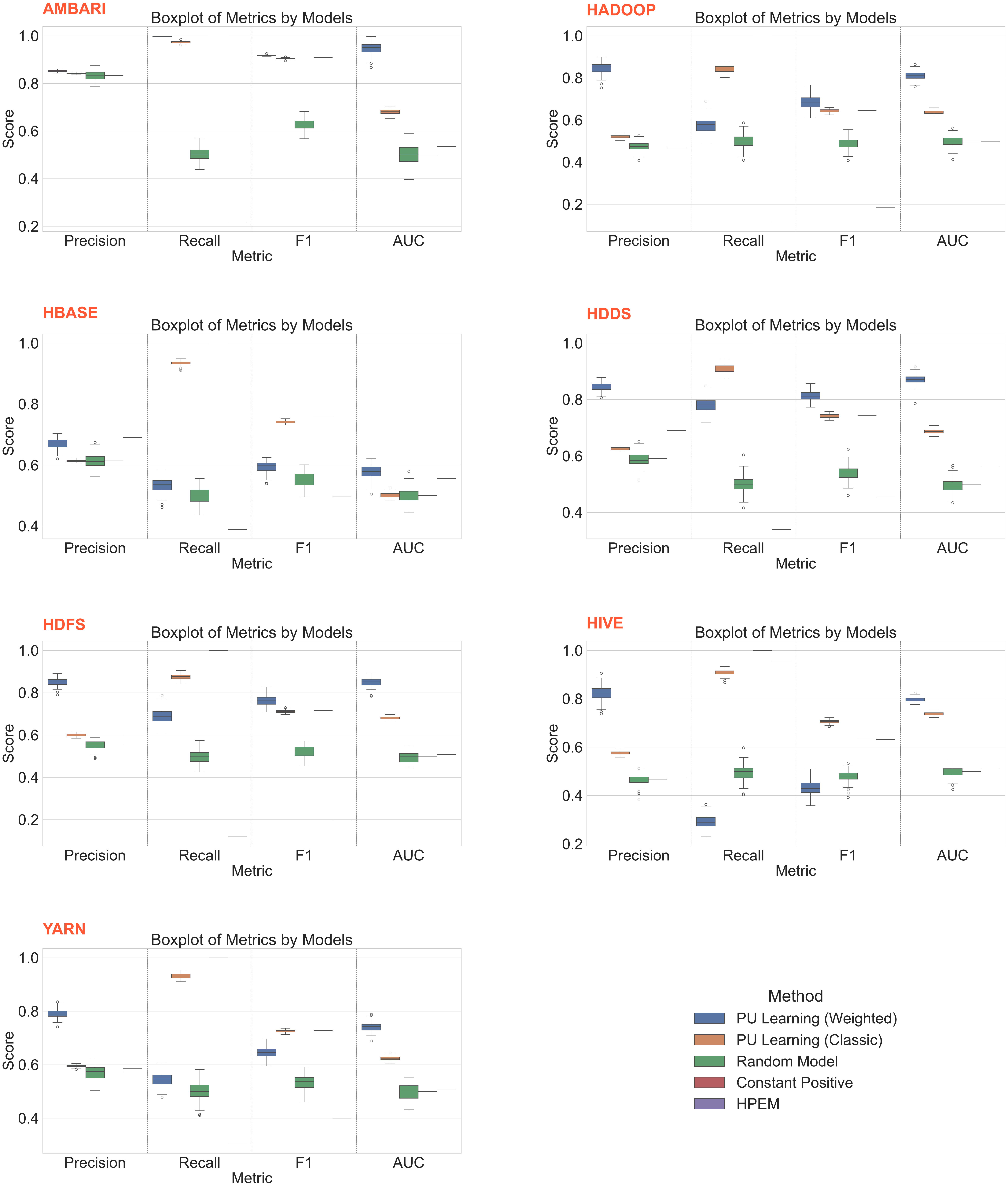}
    \caption{Performance metrics of the four selected models across the seven studied projects.}
    \label{fig:model_performance_figs}
\end{figure}
\begin{table}[ht]
    \caption{Comparison of models across different projects. 
    The first four columns represent the median values of the metrics. The subsequent four columns show the mean values with their standard deviations, which provides insights into the variability of the metrics across our 100 runs.}
    \label{tab:model_comparison}
    \centering
    \resizebox{\textwidth}{!}{%
    \begin{tabular}{llllllllll}
    \toprule
    \textbf{Project} & \textbf{Models} & \textbf{Precision} & \textbf{Recall} & \textbf{F1} & \textbf{AUC} & \textbf{Precision} \(\pm\) Std & \textbf{Recall} \(\pm\) Std & \textbf{F1} \(\pm\) Std & \textbf{AUC} \(\pm\) std \\
    \midrule
    \multirow{5}{*}{AMBARI} & PU Learning (Weighted Elkan \& Noto) & \textbf{\textcolor{red}{0.84}} & \textbf{\textcolor{red}{1.00}} & \textbf{\textcolor{red}{0.91}} & \textbf{\textcolor{red}{0.97}} & 0.84 ± 0.0034 & 1.00 ± 0.0014 & 0.91 ± 0.0021 & 0.96 ± 0.0259 \\
     & PU Learning (Classic Elkan \& Noto) & 0.83 & 0.97 & 0.89 & 0.62 & 0.83 ± 0.0025 & 0.97 ± 0.0046 & 0.89 ± 0.0025 & 0.62 ± 0.0097 \\
     & Random Model & 0.83 & 0.51 & 0.63 & 0.50 & 0.82 ± 0.0184 & 0.50 ± 0.0281 & 0.63 ± 0.0251 & 0.50 ± 0.0378 \\
     & Constant Positive Model & 0.83 & 1.00 & 0.90 & 0.50 & 0.83 ± 0.0000 & 1.00 ± 0.0000 & 0.90 ± 0.0000 & 0.50 ± 0.0000 \\
     & HPEM Approach & 0.88 & 0.22 & 0.35 & 0.54 & 0.88 ± 0.0000 & 0.22 ± 0.0000 & 0.35 ± 0.0000 & 0.54 ± 0.0000 \\
    \midrule
    \multirow{5}{*}{HADOOP} & PU Learning (Weighted Elkan \& Noto) & \textbf{\textcolor{red}{0.85}} & 0.59 & \textbf{\textcolor{red}{0.70}} & \textbf{\textcolor{red}{0.82}} & 0.85 ± 0.0309 & 0.59 ± 0.0524 & 0.70 ± 0.0402 & 0.82 ± 0.0217 \\
      & PU Learning (Classic Elkan \& Noto) & 0.52 & \textbf{\textcolor{red}{0.84}} & 0.64 & 0.64 & 0.52 ± 0.0066 & 0.84 ± 0.0168 & 0.64 ± 0.0086 & 0.64 ± 0.0063 \\
      & Random Model & 0.47 & 0.50 & 0.48 & 0.50 & 0.47 ± 0.0202 & 0.50 ± 0.0344 & 0.48 ± 0.0248 & 0.50 ± 0.0240 \\
      & Constant Positive Model & 0.47 & 1.00 & 0.64 & 0.50 & 0.47 ± 0.0000 & 1.00 ± 0.0000 & 0.64 ± 0.0000 & 0.50 ± 0.0000 \\
      & HPEM Approach & 0.47 & 0.12 & 0.19 & 0.50 & 0.47 ± 0.0000 & 0.12 ± 0.0000 & 0.19 ± 0.0000 & 0.50 ± 0.0000 \\

    \midrule
    \multirow{5}{*}{HBASE} & PU Learning (Weighted Elkan \& Noto) & \textbf{\textcolor{red}{0.70}} & 0.58 & 0.64 & \textbf{\textcolor{red}{0.63}} & 0.70 ± 0.0142 & 0.58 ± 0.0194 & 0.63 ± 0.0143 & 0.63 ± 0.0191 \\
      & PU Learning (Classic Elkan \& Noto) & 0.62 & \textbf{\textcolor{red}{0.93}} & 0.74 & 0.54 & 0.62 ± 0.0038 & 0.93 ± 0.0084 & 0.74 ± 0.0043 & 0.54 ± 0.0093 \\
      & Random Model & 0.62 & 0.50 & 0.55 & 0.50 & 0.62 ± 0.0224 & 0.50 ± 0.0297 & 0.55 ± 0.0249 & 0.50 ± 0.0270 \\
      & Constant Positive Model & 0.61 & 1.00 & \textbf{\textcolor{red}{0.76}} & 0.50 & 0.61 ± 0.0000 & 1.00 ± 0.0000 & 0.76 ± 0.0000 & 0.50 ± 0.0000 \\
      & HPEM Approach & 0.69 & 0.39 & 0.50 & 0.56 & 0.69 ± 0.0000 & 0.39 ± 0.0000 & 0.50 ± 0.0000 & 0.56 ± 0.0000 \\
    \midrule
    \multirow{5}{*}{HDDS} & PU Learning (Weighted Elkan \& Noto) & \textbf{\textcolor{red}{0.83}} & 0.80 & \textbf{\textcolor{red}{0.82}} & \textbf{\textcolor{red}{0.86}} & 0.83 ± 0.0184 & 0.79 ± 0.0339 & 0.81 ± 0.0232 & 0.86 ± 0.0321 \\
      & PU Learning (Classic Elkan \& Noto) & 0.61 & \textbf{\textcolor{red}{0.89}} & 0.73 & 0.69 & 0.61 ± 0.0059 & 0.89 ± 0.0142 & 0.73 ± 0.0074 & 0.69 ± 0.0065 \\
      & Random Model & 0.57 & 0.50 & 0.53 & 0.50 & 0.57 ± 0.0262 & 0.50 ± 0.0291 & 0.53 ± 0.0261 & 0.50 ± 0.0318 \\
      & Constant Positive Model & 0.57 & 1.00 & 0.73 & 0.50 & 0.57 ± 0.0000 & 1.00 ± 0.0000 & 0.73 ± 0.0000 & 0.50 ± 0.0000 \\
      & HPEM Approach & 0.69 & 0.34 & 0.46 & 0.56 & 0.69 ± 0.0000 & 0.34 ± 0.0000 & 0.46 ± 0.0000 & 0.56 ± 0.0000 \\
    \midrule
    \multirow{5}{*}{HDFS} & PU Learning (Weighted Elkan \& Noto) & \textbf{\textcolor{red}{0.88}} & 0.66 & \textbf{\textcolor{red}{0.75}} & \textbf{\textcolor{red}{0.85}} & 0.88 ± 0.0161 & 0.66 ± 0.0405 & 0.75 ± 0.0292 & 0.85 ± 0.0200 \\
      & PU Learning (Classic Elkan \& Noto) & 0.59 & \textbf{\textcolor{red}{0.87}} & 0.70 & 0.68 & 0.59 ± 0.0062 & 0.88 ± 0.0128 & 0.70 ± 0.0070 & 0.68 ± 0.0087 \\
      & Random Model & 0.55 & 0.50 & 0.53 & 0.50 & 0.55 ± 0.0221 & 0.50 ± 0.0293 & 0.52 ± 0.0243 & 0.50 ± 0.0242 \\
      & Constant Positive Model & 0.55 & 1.00 & 0.71 & 0.50 & 0.55 ± 0.0000 & 1.00 ± 0.0000 & 0.71 ± 0.0000 & 0.50 ± 0.0000 \\
      & HPEM Approach & 0.60 & 0.12 & 0.20 & 0.51 & 0.60 ± 0.0000 & 0.12 ± 0.0000 & 0.20 ± 0.0000 & 0.51 ± 0.0000 \\
    \midrule
    \multirow{5}{*}{HIVE} & PU Learning (Weighted Elkan \& Noto) & \textbf{\textcolor{red}{0.82}} & 0.30 & 0.44 & \textbf{\textcolor{red}{0.78}} & 0.82 ± 0.0280 & 0.30 ± 0.0275 & 0.43 ± 0.0312 & 0.78 ± 0.0100 \\
      & PU Learning (Classic Elkan \& Noto) & 0.57 & \textbf{\textcolor{red}{0.90}} & \textbf{\textcolor{red}{0.70}} & 0.73 & 0.57 ± 0.0062 & 0.90 ± 0.0111 & 0.70 ± 0.0063 & 0.73 ± 0.0068 \\
      & Random Model & 0.47 & 0.51 & 0.49 & 0.50 & 0.47 ± 0.0201 & 0.51 ± 0.0331 & 0.49 ± 0.0245 & 0.50 ± 0.0235 \\
      & Constant Positive Model & 0.46 & 1.00 & 0.63 & 0.50 & 0.46 ± 0.0000 & 1.00 ± 0.0000 & 0.63 ± 0.0000 & 0.50 ± 0.0000 \\
      & HPEM Approach & 0.47 & 0.96 & 0.63 & 0.51 & 0.47 ± 0.0000 & 0.96 ± 0.0000 & 0.63 ± 0.0000 & 0.51 ± 0.0000 \\
    \midrule
    \multirow{5}{*}{YARN} & PU Learning (Weighted Elkan \& Noto) & \textbf{\textcolor{red}{0.81}} & 0.51 & 0.63 & \textbf{\textcolor{red}{0.74}} & 0.80 ± 0.0200 & 0.51 ± 0.0311 & 0.62 ± 0.0265 & 0.74 ± 0.0170 \\
      & PU Learning (Classic Elkan \& Noto) & 0.58 & \textbf{\textcolor{red}{0.92}} & 0.71 & 0.62 & 0.58 ± 0.0042 & 0.92 ± 0.0105 & 0.71 ± 0.0050 & 0.62 ± 0.0084 \\
      & Random Model & 0.56 & 0.50 & 0.53 & 0.50 & 0.56 ± 0.0219 & 0.50 ± 0.0285 & 0.53 ± 0.0239 & 0.50 ± 0.0257 \\
      & Constant Positive Model & 0.56 & 1.00 & \textbf{\textcolor{red}{0.72}} & 0.50 & 0.56 ± 0.0000 & 1.00 ± 0.0000 & 0.72 ± 0.0000 & 0.50 ± 0.0000 \\
      & HPEM Approach & 0.59 & 0.30 & 0.40 & 0.51 & 0.59 ± 0.0000 & 0.30 ± 0.0000 & 0.40 ± 0.0000 & 0.51 ± 0.0000 \\
    \bottomrule
    \end{tabular}%
    }
\end{table}

\Fig~\ref{fig:model_performance_figs} shows the performance boxplots of our models across each project, evaluated using four key metrics: Precision, Recall, F1, and AUC. Each subplot represents the distribution of scores for our selected four different models, which showcases the variations in model effectiveness.
\Tbl~\ref{tab:model_comparison} displays the details of the performance in each model (by median and mean\(\pm\)std values) for our studied projects. 
We observe that our classic and weighted Elkan \& Noto model consistently outperforms baseline models (Random Model, Constant Positive Model and HPEM approach) across most metrics (\ie{} Precision, F1, AUC.).
Specifically, the AUC values for the weighted Elkan \& Noto model range from 0.62 (at HIVE) to 0.97 (at AMBARI), showcasing its ability to predict the unrelated build failures among a set of unlabeled data~\cite{ccorbaciouglu2023receiver}.
In the AMBARI project, the AUC value for the weighted Elkan \& Noto model can reach as high as 0.97.
One possible reason is that AMBARI has a highly imbalanced class distribution, with positive instances comprising 82.55\% of the data. Hence, the AUC value can benefit from the weighted adaptation of PU learning, optimizing the model’s focus on more significant positive samples.
In projects like HDFS and YARN, the AUC scores of the weighted Elkan \& Noto model maintain high median values (The value of AUC is 0.85 at HDFS and 0.81 at YARN) but are slightly lower compared to AMBARI and HADOOP. However, the weighted Elkan \& Noto model significantly outperforms the baseline models.
The baseline models consistently exhibit lower AUC, Precision, and F1 scores, underscoring the predictive capabilities of our PU Learning models in identifying unrelated build failures. 
This demonstrates that our models are effective tools for detecting.

Overall, our results suggest that automatically predicting unrelated build failures for developers is feasible. Based on the predictions, developers can obtain additional information, helping their decision-making on whether they should proceed with the merge. An indication of the likelihood that a failure is unrelated can also reduce efforts on investigating the cause for the build breakage, since the build failure is indicated to be unrelated to the current push.

{\fontsize{8}{11}\selectfont

\begin{table}[!h]
\centering
\caption{Feature Importance Results (mean \textpm std values) from PU (Weighted) Learning}
\resizebox{0.95\linewidth}{!}{
\begin{tabular}{p{1.8cm}|p{1.8cm}|p{1.8cm}|p{1.8cm}|p{1.8cm}|p{1.8cm}|p{1.8cm}|p{1.8cm}}
\toprule
\diagbox[width=8em, dir=NW,font=\small]{\textbf{Feature}}{\textbf{Project}} & \textbf{AMBARI} & \textbf{HADOOP} & \textbf{HBASE} & \textbf{HDDS} & \textbf{HDFS} & \textbf{HIVE} & \textbf{YARN} \\
\midrule
CI Latency & 0.3593 \textpm 0.0021 \highlight{\textbf{(Top 1)}} & 0.3711 \textpm 0.0016 \highlight{\textbf{(Top 1)}} & 0.3042 \textpm 0.0016 \highlight{\textbf{(Top 1)}} & 0.2244 \textpm 0.002 \highlight{\textbf{(Top 1)}} & 0.2814 \textpm 0.0018 \highlight{\textbf{(Top 1)}} & 0.1437 \textpm 0.0015 \highlight{\textbf{(Top 3)}} & 0.2342 \textpm 0.0015 \highlight{\textbf{(Top 1)}} \\
Is Shared Same Emsg & - & - & - & - & - & 0.1514 \textpm 0.0015 \highlight{\textbf{(Top 2)}} & - \\
Modified Source Code Files & - & - & - & - & - & - & 0.1406 \textpm 0.0012 \highlight{\textbf{(Top 3)}} \\
Number of Parallel Issues & 0.2895 \textpm 0.002 \highlight{\textbf{(Top 2)}} & - & - & - & - & - & - \\
Number of Prior Comments & - & 0.2095 \textpm 0.0015 \highlight{\textbf{(Top 2)}} & 0.1966 \textpm 0.0015 \highlight{\textbf{(Top 3)}} & 0.1575 \textpm 0.0014 \highlight{\textbf{(Top 3)}} & 0.2173 \textpm 0.0016 \highlight{\textbf{(Top 2)}} & - & 0.172 \textpm 0.0014 \highlight{\textbf{(Top 2)}} \\
Number of Similar Failures & 0.119 \textpm 0.0013 \highlight{\textbf{(Top 3)}} & 0.0872 \textpm 0.0011 \highlight{\textbf{(Top 3)}} & 0.1986 \textpm 0.0015 \highlight{\textbf{(Top 2)}} & 0.1691 \textpm 0.0014 \highlight{\textbf{(Top 2)}} & 0.1545 \textpm 0.0013 \highlight{\textbf{(Top 3)}} & 0.1892 \textpm 0.002 \highlight{\textbf{(Top 1)}} & - \\
\bottomrule
\end{tabular}
}
\label{tab:fi_for_model}
\end{table}
}

\subsection{\textbf{Generalization of PU-Learning Models}}

\begin{table*}[ht]
\centering
\caption{Cross-Project Validation Results}
\label{tab:cross_project_validation}
\begin{tabular}{lccccc}
\toprule
\textbf{Test Project} & \textbf{Model} & \textbf{Precision} & \textbf{Recall} & \textbf{F1} & \textbf{AUC} \\
\midrule
\multirow{2}{*}{AMBARI} 
 & Classic Elkan \& Noto  & 0.583 $\pm$ 0.002 & \textbf{0.884 $\pm$ 0.004} & 0.702 $\pm$ 0.002 & 0.647 $\pm$ 0.003 \\
 & Weighted Elkan \& Noto & \textbf{0.781 $\pm$ 0.027} & 0.809 $\pm$ 0.038 & \textbf{0.794 $\pm$ 0.027} & \textbf{0.842 $\pm$ 0.035} \\
\midrule
\multirow{2}{*}{HADOOP} 
 & Classic Elkan \& Noto  & 0.641 $\pm$ 0.002 & 0.902 $\pm$ 0.005 & 0.749 $\pm$ 0.002 & 0.684 $\pm$ 0.003 \\
 & Weighted Elkan \& Noto & \textbf{0.816 $\pm$ 0.020} & \textbf{0.967 $\pm$ 0.011} & \textbf{0.885 $\pm$ 0.013} & \textbf{0.962 $\pm$ 0.013} \\
\midrule
\multirow{2}{*}{HBASE} 
 & Classic Elkan \& Noto  & 0.631 $\pm$ 0.003 & 0.882 $\pm$ 0.005 & 0.735 $\pm$ 0.003 & 0.704 $\pm$ 0.003 \\
 & Weighted Elkan \& Noto & \textbf{0.819 $\pm$ 0.021} & \textbf{0.970 $\pm$ 0.010} & \textbf{0.888 $\pm$ 0.013} & \textbf{0.969 $\pm$ 0.011} \\
\midrule
\multirow{2}{*}{HDDS} 
 & Classic Elkan \& Noto  & 0.628 $\pm$ 0.002 & 0.892 $\pm$ 0.005 & 0.737 $\pm$ 0.003 & 0.683 $\pm$ 0.003 \\
 & Weighted Elkan \& Noto & \textbf{0.801 $\pm$ 0.020} & \textbf{0.958 $\pm$ 0.011} & \textbf{0.873 $\pm$ 0.013} & \textbf{0.954 $\pm$ 0.014} \\
\midrule
\multirow{2}{*}{HDFS} 
 & Classic Elkan \& Noto  & 0.634 $\pm$ 0.003 & 0.894 $\pm$ 0.005 & 0.742 $\pm$ 0.003 & 0.687 $\pm$ 0.003 \\
 & Weighted Elkan \& Noto & \textbf{0.812 $\pm$ 0.020} & \textbf{0.971 $\pm$ 0.008} & \textbf{0.884 $\pm$ 0.012} & \textbf{0.969 $\pm$ 0.009} \\
\midrule
\multirow{2}{*}{HIVE} 
 & Classic Elkan \& Noto  & 0.638 $\pm$ 0.003 & 0.894 $\pm$ 0.005 & 0.745 $\pm$ 0.003 & 0.673 $\pm$ 0.003 \\
 & Weighted Elkan \& Noto & \textbf{0.810 $\pm$ 0.015} & \textbf{0.975 $\pm$ 0.007} & \textbf{0.885 $\pm$ 0.010} & \textbf{0.968 $\pm$ 0.009} \\
\midrule
\multirow{2}{*}{YARN} 
 & Classic Elkan \& Noto  & 0.626 $\pm$ 0.002 & 0.886 $\pm$ 0.005 & 0.734 $\pm$ 0.003 & 0.684 $\pm$ 0.003 \\
 & Weighted Elkan \& Noto & \textbf{0.823 $\pm$ 0.020} & \textbf{0.961 $\pm$ 0.012} & \textbf{0.886 $\pm$ 0.012} & \textbf{0.965 $\pm$ 0.009} \\
\bottomrule
\end{tabular}
\end{table*}

Table~\ref{tab:cross_project_validation} presents the results of cross-project validation, where models trained on data from six projects and tested on the held-out seventh project. Each row reports the performance of our model on the held-out project, with the model trained on the remaining six. We compare the performance of two PU learning models, the Classic Elkan \& Noto and Weighted Elkan \& Noto, using four evaluation metrics, including precision, recall, F1-score, and AUC. Across all seven projects, the Weighted Elkan \& Noto model consistently outperforms the Classic Elkan \& Noto model on all four metrics. The weighted variant achieves substantially higher precision, with improvements ranging from ~13\% to ~24\%, indicating a stronger ability to avoid false positives (i.e., misclassifying related failures as unrelated). For instance, on the AMBARI project, precision improves from 0.583 to 0.781, and on HDFS, from 0.634 to 0.812. In terms of recall, both models perform strongly, with the weighted model maintaining or improving upon already high recall values ($\geq$
0.958 across projects except AMBARI). This suggests the model is effective at identifying truly unrelated failures. The F1-score, which balances precision and recall, also shows substantial gains with the weighted model (\eg improving from 0.735 to 0.888 in HBASE and from 0.702 to 0.794 in AMBARI), highlighting the overall robustness of the approach. Similarly, the AUC scores increase significantly with the Weighted Elkan \& Noto PU learning , demonstrating stronger discriminative power of the model (\eg HADOOP improves from 0.684 to 0.962, HDFS from 0.687 to 0.969).

Overall, the results demonstrate that our proposed Weighted Elkan \& Noto PU learning model generalizes effectively across different projects. The weighted variant consistently outperforms the classic model in terms of precision, F1, and AUC metrics, establishing itself as a robust and reliable method for identifying unrelated build failures. Furthermore, the model maintains strong performance even on projects with limited historical data, highlighting its applicability in low-history environments and its potential for real-world continuous integration systems.

\subsection{\textbf{Importance of Features to Predict Unrelated Build Failures}}
As shown in \Tbl \ref{tab:fi_for_model}, the features \texttt{CI latency}, \texttt{Number of Similar Failures} and \texttt{Number of Prior Comments} significantly impact the prediction of unrelated CI failures across our study cases. 

\texttt{CI Latency} represents the time elapsed from code patch submission to the push triggering the build. The rationale behind this feature is that if a submitted code patch is quickly pushed, it is more likely that the build failure is related. Conversely, if there is a delay in pushing the patch, the resulting build failure may be unrelated due to other pushes in the meantime. The \texttt{CI Latency} feature is consistently among the top 3 most important features in our seven studied projects.

\texttt{Number of Similar Failures} indicates whether a CI failure shares a similar error message with a past build failure, meaning that if a past failure presents a similar error message to the current failure, it may indicate that the present failure is a repetition of a previous, and unrelated, failure (i.e., the cascading failures discussed by Ghaleb et al.~\cite{ghaleb2019empirical}). This factor is \ among the top 3 most important features in  \textsc{HIVE}(Top 1), \textsc{HBASE}(Top 2), \textsc{HDDS}(Top 2) \textsc{AMBARI}(Top 3), \textsc{HADOOP} (Top 3) and \textsc{HDFS}(Top 3).

The \texttt{Number of Prior Comments} also emerges as a top 3 important feature in the five projects: \textsc{HADOOP}, \textsc{HBASE}, \textsc{HDDS}, \textsc{HDFS} and \textsc{YARN}. This feature helps gauge the historical discussion around an issue, which can significantly indicate whether a build failure is unrelated to the present push. For instance, if an issue requires extensive discussion, it may suggest complexity and connections to other issues, resulting in build failures that are unrelated to the current push, but related to other issues.

Overall, the important features of our model can help developers reason about whether the build failure is likely related or unrelated to their code push. By regularly monitoring these features and taking proactive actions, teams can iteratively mitigate the negative impact of unrelated failures, enhancing their build processes and software quality.

%% file: src/tables/rq1_tbl2.tex
{\fontsize{8}{11}\selectfont
\begin{longtable}[!h]{m{3cm}m{7cm}m{0.8cm}}
\caption{The Explanation of Inductive Themes from Document Analysis} \label{results:rq1_tbl2} \\
\toprule

\textbf{Theme} & \textbf{Description} & \textbf{Ratio (Amount)} \\
\midrule
\endfirsthead

\multicolumn{3}{c}%
{{\tablename\ \thetable{} -- continued from previous page}} \\
\midrule
\textbf{Theme} & \textbf{Description} & \textbf{Ratio (Amount)} \\
\midrule
\endhead

\multicolumn{3}{r}{{Continued on next page}} \\
\endfoot

\midrule
\endlastfoot
\textit{Concurrent Commits} & Refers to the occurrence of parallel commits in a CI environment. It can generate unpredictable errors in the codebase due to the unforeseen interactions between the changes. & \textit{3.24\% (12)} 
\\\midrule
\textit{Configuration-related Issues} &  Refers to failures or problems encountered by a QA bot or any other system due to incorrect or faulty configuration settings. These issues are unrelated to the code pushes or topics addressed in the current issue report. & \textit{2.43\% (9)} 
\\\midrule
\textit{External Interference} & Failures that occur when dependencies or external resources, such as third-party libraries or package registries, are updated, unavailable, or behave unexpectedly. For example, a build fails because a new version of a JSON-parsing library introduces a breaking change, even though the current code patch did not modify anything related to that library. & 19.73\% (73)
\\
\midrule
\textit{Environment Impact} & Refers to failures or issues that occur in a software development or testing process due to the environmental factors, such as Jenkins, Jira, timeout during the CI process, or already used addresses, and so forth & \textit{2.43\% (9)} \\ 
\midrule
\textit{Non-reproducible} & Failures that pass when re-running the same commit on a fresh environment, often due to transient infrastructure problems (e.g., disk corruption, network timeouts, clock skew, container cold starts). For instance, a build passes locally and on a re-run in CI, but would fail on another environment due to different configurations that are not easily apparent initially. & \textit{17.03\% (64)} \\ \midrule
\textit{Unrelated Tests} & Refers to errors or failures that exist in testing classes, which are unrelated to the changes introduced in the current push. & \textit{20.00\% (74)} \\ \midrule
\textit{Unspecified by the Developer} & Refers to any build failures that are reported to be unrelated to the current push but were not discussed in detail, so the reason for deeming a build failure as unrelated could not be derived. & \textit{34.86\% (129)} \\
\end{longtable}
}

%% file: src/tables/rq1_tbl3.tex
\setlength{\tabcolsep}{1.5pt}  

\begin{table}[htbp]
\centering
\caption{Theme Distribution by Project}
\label{tab:theme-distribution}
\begin{minipage}[t]{0.48\textwidth}
\centering
\scriptsize
\begin{tabular}{llrr}
\toprule
\textbf{Project} & \textbf{Theme} & \textbf{Count} & \textbf{\%} \\
\midrule
\multirow{2}{*}{AMBARI}
  & Unspecified by the Developer                      & 3 & 75.00 \\
  & Configuration-related Issues & 1 & 25.00 \\
\midrule
\multirow{7}{*}{HADOOP}
  & Unspecified by the Developer                      & 13 & 33.33 \\
  & External Interference        & 9  & 23.08 \\
  & Non-reproducible             & 8  & 20.51 \\
  & Unrelated Tests              & 5  & 12.82 \\
  & Environment Impact           & 2  & 5.13 \\
  & Concurrent Commits           & 1  & 2.56 \\
  & Configuration-related Issues & 1  & 2.56 \\
\midrule
\multirow{7}{*}{HBASE}
  & Unspecified by the Developer                      & 21 & 38.18 \\
  & Non-reproducible             & 15 & 27.27 \\
  & External Interference        & 9  & 16.36 \\
  & Unrelated Tests              & 6  & 10.91 \\
  & Configuration-related Issues & 2  & 3.64 \\
  & Concurrent Commits           & 1  & 1.82 \\
  & Environment Impact           & 1  & 1.82 \\
\midrule
\multirow{6}{*}{HDDS}
  & Unspecified by the Developer                      & 7  & 41.18 \\
  & External Interference        & 4  & 23.53 \\
  & Environment Impact           & 2  & 11.76 \\
  & Unrelated Tests              & 2  & 11.76 \\
  & Concurrent Commits           & 1  & 5.88 \\
  & Non-reproducible             & 1  & 5.88 \\
\bottomrule
\end{tabular}
\end{minipage}
\hfill
\begin{minipage}[t]{0.48\textwidth}
\centering
\scriptsize
\begin{tabular}{llrr}
\toprule
\textbf{Project} & \textbf{Theme} & \textbf{Count} & \textbf{\%} \\
\midrule
\multirow{7}{*}{HDFS}
  & Unspecified by the Developer                      & 41 & 36.94 \\
  & External Interference        & 27 & 24.32 \\
  & Non-reproducible             & 22 & 19.82 \\
  & Unrelated Tests              & 13 & 11.71 \\
  & Concurrent Commits           & 4  & 3.60 \\
  & Environment Impact           & 3  & 2.70 \\
  & Configuration-related Issues & 1  & 0.90 \\
\midrule
\multirow{7}{*}{HIVE}
  & Unrelated Tests              & 39 & 41.94 \\
  & Unspecified by the Developer                      & 31 & 33.33 \\
  & External Interference        & 12 & 12.90 \\
  & Non-reproducible             & 7  & 7.53 \\
  & Concurrent Commits           & 2  & 2.15 \\
  & Configuration-related Issues & 1  & 1.08 \\
  & Environment Impact           & 1  & 1.08 \\
\midrule
\multirow{6}{*}{YARN}
  & Unspecified by the Developer                      & 13 & 25.49 \\
  & External Interference        & 12 & 23.53 \\
  & Non-reproducible             & 11 & 21.57 \\
  & Unrelated Tests              & 9  & 17.65 \\
  & Concurrent Commits           & 3  & 5.88 \\
  & Configuration-related Issues & 3  & 5.88 \\
\bottomrule
\end{tabular}
\end{minipage}
\end{table}

%% file: src/relatedwork.tex
In this section, we examine research that has been conducted on Continuous Integration and Continuous Delivery, with a focus on empirical studies. 
We also consider related work on the problem of flaky tests occurring during the CI/CD process as flaky tests are loosely related to the concept of unrelated build failures. 

\subsection{Empirical Studies on CI Failures Categorization and Improvement}

Continuous integration has become an important part of software development. 
Continuous integration can improve the code quality and the delivery ability of the engineering team \cite{fowler2006continuous,meyer2014continuous,shahin2017continuous,arachchi2018continuous,bernardo2018studying}.
Hence, the research on CI has attracted people's attention.

Vassallo \etal \cite{vassallo2020principle} researched CI practices, identified anti-patterns, and developed Bart, CD-Linter, and CI-Odor tools. 
The empirical studies showed the solutions were effective in promoting principled CI practices and preventing decay over time. CI-Odor was validated and found useful by developers.
Rausch \etal \cite{rausch2017empirical} analyzed CI build failures in 14 Java projects, identifying 14 error categories (like test failures, and compile errors) and emphasized the need to address flaky tests and maintain healthy build systems for effective CI.

Because CI/CD processes are essential for software development, researchers are interested in ways to make them better.
To increase the reliability and effectiveness of Continuous Integration and Continuous Deployment (CI/CD) procedures, Gallaba \etal \cite{gallaba2019improving} conducted empirical research and analyzed the noise and heterogeneity in CI/CD outcome data, discovered anti-patterns in CI/CD requirements, and raised \textit{Hansel and Gretel} as tools and inference-based build acceleration to quicken CI builds. 
Shi \etal \cite{shi2019understanding} compared module- and class-level regression test selection (RTS) techniques in a cloud-based CI environment and developed a hybrid RTS technique. 
RTS reduced testing time and helped avoid false alarms from flaky tests, providing benefits to developers in the CI environment.

To make the CI process more effective  
Elbaum \etal \cite{elbaum2014techniques} studied the usage of Automated Static Code Analysis Tools (ASCATs) within Continuous Integration (CI) in projects using Travis CI. They provided an effective and efficient adoption of ASCATs in CI, including excluding unnecessary checks, properly configuring tools according to coding guidelines, and documenting decisions to suppress warnings.
Knauss \etal \cite{knauss2015supporting} evaluated a method, which is based
on analysis of correlations between test-case failures and source
code changes, for selecting a suitable set of functional regression tests on the system level, and found an optimal test suite to execute before the integration.
Hassan \cite{hassan2019tackling} developed an approach called \textit{HireBuild} that can automatically fix build errors involving build scripts.
Reinforcement learning techniques, for example, have proved essential in enhancing the CI process.
Yang \etal \cite{yang2020systematic} conducted a systematic study of the reward function and applied reward strategy in CI testing and sought to improve CI testing efficiency.
Similarly, two novel reward functions were proposed by Li \etal \cite{li2021weighted}, which focused on the impact of failure position in the test case history execution sequence, which is the Average Position Exponential Weight (APEW) reward function and the Average Position Quadratic Weight (APQW) reward function, respectively.

These studies focused on CI build failures of all types, including those caused by code pushes or not.
Instead of focusing on detecting general CI failures, our focus is on identifying build failures that are unrelated to the most recent push. Our approach can save developers from efforts on checking whether a build failure is related to their current push. Our work complements previous research, since a development team can benefit from multiple approaches when managing their CI pipelines.
\subsection{Studies on the Flaky Tests}

In our work, we investigate {\bfseries legitimate} failures that are unrelated to the current push. However, a related problem is that of flaky tests, which are tests that produce inconsistent results, posing a significant challenge in software testing. There has been an increased focus on the issue of test flakiness in recent years, with a number of studies exploring the causes and effects of flaky tests in both open-source and proprietary software. 

Luo \etal \cite{luo2014empirical}'s research on flaky tests can be traced back to their study of 51 Apache projects, in which they classified the root causes of these tests. The results showed that flaky tests undermine regression testing, caused by race conditions, external dependencies, and stateful tests.
This study provided insights and implications that can guide future research on the topic of avoiding flaky tests.
Eck \etal \cite{eck2019understanding} categorized 200 tests and polled 121 developers online to determine how developers perceive flaky tests. They discovered four new flaky test categories, including the effects of flakiness on resource allocation and software reliability, as well as difficulties in recreating and spotting flaky behavior.
Ahmad \etal \cite{ahmad2021empirical} conducted a survey with developers to understand their perceptions of test flakiness, including how they define flaky tests and the factors that contribute to their presence. The study found that several key factors, including software product quality and the quality of the test suite, are affected by test flakiness.
Previous research \cite{liblit2005scalable,palomba2017notice,thorve2018empirical} revealed that concurrency is the main reason for flaky tests, therefore detecting and repairing concurrency can partially alleviate flaky tests. Silva \etal~\cite{silva2020shake} proposed SHAKER, a lightweight technology, which is more effective and resource-saving than RERUN for detecting the flaky test caused by concurrency.

Flaky build failures can also be caused by configuration and infrastructure aspects.
\textsc{VeriCI}, a method by Santolucito \etal \cite{santolucito2022learning} for localizing CI configuration issues at the code level, takes advantage of commit and build history in CI settings to forecast build failures and pinpoint root causes. \textsc{VeriCI} was introduced to help improve the CI process and avoid CI failures which are not triggered by code modifications.

Adriaan \etal~\cite{labuschagne2017measuring} identified the flaky tests from Travis and found that 18\% of test suite executions fail and that 13\% of these failures are flaky. They provided an approach to identify the flaky tests based on the build status history.
Doriane \etal~\cite{olewicki2022towards} tried to detect flaky tests based on textual similarity to build logs of prior brown builds (a build failure that changes to success on at least one build rerun without changing the build setup or
source code) and the model achieved the F-score value at 53\% on the studied projects.
Johannes \etal~\cite{lampel2021life} drew attentions on intermittent failures, and proposed classification models using job telemetry data to diagnose failure patterns. Their approach achieves precision scores of 73\%.

Unrelated failures are a different problem from flaky tests. Unrelated failures are indeed legitimate and not non-deterministic like flaky tests. However, although certain CI failures are legitimate they may not be related to the push of a certain developer, which is a common scenario in intensive software development projects. The relatedness of a failure to code pushes can impact the decision to merge a change.

\subsection{CI Build Outcome Prediction}

Build failures in Continuous Integration (CI) systems often occur during the software development process. Previous studies have aimed to develop predictive models to identify when a build is likely to fail in advance and mitigate failures. Jin and Servant~\cite{jin2020cost} introduced a cost-effective approach to CI by predicting build failures and selectively skipping builds that are unlikely to succeed, thus reducing unnecessary resource consumption and accelerating feedback loops. Chen \etal~\cite{chen2020buildfast} proposed a history-aware build outcome prediction model that leverages historical build data to predict future build outcomes, aiming to mitigate potential build failures. Saidani \etal~\cite{saidani2020prediction} developed a multi-objective genetic programming approach to predict build failures, addressing the imbalance of build outcomes and enhancing the build process. Machalica \etal~\cite{machalica2019predictive} proposed a predictive test selection strategy that selects a subset of tests to perform for each build change, reducing infrastructure costs while maintaining high fault detection rates. Although these studies focus on predicting build failures based on various factors such as historical data, developer changes, and test outcomes, our research specifically targets ``unrelated build failures" that occur due to factors not directly related to the recent code changes, such as misconfigurations, overlooked dependencies, or unintended interactions between components. Our approach uncovers unrelated build failure patterns that traditional models miss, and aims to identify and mitigate unrelated failures by analyzing issue reports fileds, and their interdependencies, providing a more targeted solution to a specific class of build failures.

Our main goal is to identify build failures that are unrelated to the recent push, which may confuse developers in their decision-making process because they invest time investigating how to fix a failure, which can lead them to perceive that such a failure is not related to their push. Upon discovering that failures are unrelated, developers may choose to integrate their changes if the team assesses that it is safe to do so. Alternatively, they could at least understand the actual scope of the failure and target its root cause without thinking that the failure is related to their current push. As such, our work is substantially different from previous CI build outcome prediction research.

One could argue that unrelated failures would be impossible to occur when proper CI practices are in place, \ie only merging code upon passing builds. However, in large codebases with multiple developers working concurrently, such as the projects we study and those examined by Ghaleb \etal~\cite{ghaleb2019empirical}, one developer's change can accidentally impact another's work, potentially causing future build failures that appear to be unrelated to original contributions. Furthermore, as noted by Felidré \etal~\cite{felidre2019continuous}, projects employing CI practices do not always adhere to the highest standards, which could lead to increased occurrence of unrelated build failures. Therefore, our research is valuable for projects utilizing CI, as it acknowledges that achieving perfection in adhering to these practices may not always be feasible in real-world scenarios.

%% file: src/discussion.tex
This section explores the implications of our findings for developers and researchers, particularly in making decisions about whether a build failure is attributable to the current code push.

\subsection{\textbf{Implications for developers}}
\textbf{Building upon the results of our study about leveraging the PU learning models into the development workflow can enhance the efficiency of debugging.}
Applying our prediction model by PU learning in the development process can give developers an indication of whether a build failure is related to the current code push, which reduces the time spent on manual investigation by developers and enables a faster resolution for developers on continuous integration build issues.
By accurately determining the root cause of build failures, developers can allocate their time and resources more effectively and focus on issues within their control rather than unrelated external factors.
\\
\\
\textbf{Developers should be aware of the recurring patterns of unrelated build failures.}
Developers can use the insights from our PU learning model to proactively address recurring patterns (\ie our found feature importance of PU learning models in Section~\ref{methodology}) of unrelated build failures. 
For example, the long CI latency, where developers delay the code build after the code was pushed.
To address the long CI latency, developers can be advised to trigger the build of the code push as soon as possible, minimizing the occurrence of unrelated build failure warnings. 
This proactive approach not only reduces unnecessary noise in the CI workflow but also ensures that failure warnings are more relevant and timely to the current code pushes. As a result, the overall stability and reliability of the software development life cycle are enhanced, since the system can more effectively differentiate between valid issues and false alarms, leading to more efficient troubleshooting and quicker resolutions.

\subsection{\textbf{Implications for researchers}}
\textbf{Researchers should pay closer attention to unrelated failures in code pushes.}
Based on our findings in Section~\ref{motivating:example}, which indicate that the median time spent identifying unrelated build failures exceeds 4 hours in the studied projects, we recommend that researchers delve deeper into uncovering insights about failures unrelated to code pushes.
Although prior research has focused on flaky tests occurring during the test workflow, our results in Section~\ref{results} reveal that it is not sufficient to look at only the flaky test.
By studying such unrelated failures, researchers can gain deeper insights into the factors influencing how different build failures impact the efficiency of the software development process.
\\
\\
\textbf{Researchers should explore broader applications of PU learning across diverse software engineering domains.}
Our work reveals the potential of PU learning in addressing challenges where labels about whether a build failure is related to current code push are not reliable. 
Missing labels are a common challenge in real-world applications, particularly when researchers aim to apply machine learning for classification tasks.
Our work demonstrates that PU learning can serve as a feasible and promising methodology for addressing other Software Engineering challenges, such as detecting rare bugs, identifying anomalies in log analysis, or uncovering under-documented code segments.

%% file: src/threatsToValidity.tex
In this section, we discuss the threats to validity of our work.

\textbf{Construct validity} refers to how our inferences are supported by the evidence. 

Throughout our methodology, there are several stages where we encounter threats to validity. 
The heuristic-based approach we employ to identify potentially unrelated build failures results in false positives. 
To further evaluate the accuracy of our heuristic, we compared its labels against manually validated labels from dataset \(Q\) (1,179 manually confirmed unrelated failures). 
We retrieved 335 potential unrelated failures identified by our heuristic and found that it is conservative, tending to underestimate rather than overestimate unrelated failures.
Moreover, the false positive rate is low, only 26 out of 335 heuristic positives (\textless 8\%) 
These findings suggest that our reported prevalence of 13.33\% (Section 2) should be interpreted cautiously as a lower bound, as the true prevalence of unrelated failures is likely higher. 
This limitation directly motivates our use of PU learning, which is designed to handle incomplete positive labeling by learning from both confirmed positives and unlabeled data.

Another threat is related to the assumption of ``\textit{selected completely at random}'' (\textbf{SCAR}) when building our PU learning models. For instance, if the SCAR assumption does not hold, the evaluations performed in our models would likely not represent the actual performance in the latent positive instances hidden in the unlabeled population. However, we mitigate this problem by expanding our truly unrelated failures dataset to 700 instances, which is enough to manifest a statistically representative sample of 95\% confidence level with 4\% margin of error. As such, we are confident that our sample used to build our models represents of the distribution of unrelated build failures and, therefore, the SCAR assumption is valid.

Additionally, the metric we use to measure the delay or time-to-recognition of failures, based on timestamps from issue tracking systems, may not fully capture the actual developer effort or the time actively spent diagnosing the failures. Instead, it reflects the elapsed time recorded in the repository. This limitation is inherent in empirical software engineering research that relies on archival data from software repositories, and while it may affect the precision of inferences about developer engagement, it does not undermine the overall validity of our findings.

\textbf{External validity} addresses the extent to which our findings can be applied to different contexts, including other software projects. While our paper studies only 7 projects, our main objective was not to achieve the utmost generalizability. Rather, our main objective was to develop an automated approach for identifying unrelated build failures. For this purpose, we use as case studies projects where Continuous Integration (CI) practices are well-established. Through our approach, we achieved promising results, with models exhibiting AUC values from 0.62 to 0.97 in the studied projects.
We believe that our methodology can be extended to different projects, provided the necessary features can be computed. However, adaptations might be necessary for varying project types, such as the calculation of additional metrics.

\textbf{Internal validity} refers to the degree of confidence or certainty with which our study can establish a causal relationship between the independent variable(s) and the dependent variable(s).
To mitigate the risk of misclassifying relevant failures as unrelated, we apply a conservative high decision threshold that prioritizes precision. In addition, rather than hiding failures predicted as unrelated, the model highlights them on the CI dashboard along with a confidence score. This design provides developers with additional information to support their debugging process, allowing them to review and override predictions as needed to ensure that no critical issues are overlooked.

Another point is, although certain features are derived from JIRA, the proposed approach can be applied to other CI/CD environments where JIRA is not used. Equivalent features can be obtained from GitHub or other platforms; for example, cross-organization commit patterns or repository dependencies may substitute for “Is Cross Project.” Consequently, the approach is generalizable to diverse development environments, including GitHub Actions, CircleCI, and TravisCI.

Also, to reduce the risk of overfitting, we conduct PU-Learning with 10-fold cross-validation for 100 iterations. The time-wise and iterative nature of the process allows for repeated model training, ensuring that the model learns from a variety of data selections, yielding a more reliable and representative outcome. 

The use of PU-Learning also helps address the problem with the presence of unbalanced data, where positive instances (unrelated failures) are much rarer than negative instances (related failures).

%% file: src/conclusion.tex
In this research paper, we study {unrelated} build failures, which are not caused by the pushes that triggered them. We characterize and predict the unrelated build failures of the \textsc{AMBARI}, \textsc{HBASE}, \textsc{HDFS}, \textsc{HDDS}, \textsc{HADOOP}, \textsc{YARN}, and \textsc{HIVE} projects. We observe that unrelated build failures take a median of 4 hours to be identified as such, becoming a burden on developers' time.

Our document analysis on the reasons behind unrelated build failures reveals that \UnrelatedTestingClasses~ and \ExternalInterference~ are among the main reasons why developers consider certain build failures unrelated. 
To predict whether a build failure is unrelated, we propose the use of a semi-supervised learning approach based on positive and unlabeled instances (PU Learning).
Our PU-learning models thus not only achieve AUC values from $0.63 \pm 0.02$ to $0.97 \pm 0.03$ but also demonstrate precision between $0.70 \pm 0.01$ and $0.88 \pm 0.01$, recall between $0.30 \pm 0.03$ and $1.00 \pm 0.00$, and F$_1$-scores between $0.44 \pm 0.03$ and $0.91 \pm 0.00$, outperforming all baselines and showing that our semi-supervised framework can efficiently detect unrelated build failures.
Our prediction model aims to give developers an index of decision-making. 
For instance, the developers can know the probability of whether a build failure is (un)related to a code push, and then can lower the priority in investigating the issues in code based on the prediction result. 
Finally, we identify (i) the time taken from a submitted patch to the build-triggering push (CI latency); (ii) build failures sharing similar error messages with recent failures; and (iii) the number of comments preceding the build failure all efficient indicators for identifying unrelated build failures. 

Our research shows promising results in the identification of unrelated build failures. Our next step is to apply our approach in industry settings to gather feedback about its efficiency and opportunities for improvements.

%% file: src/appendix/appendix_2.tex
\section*{Explanation of Our Feature Engineering}
\label{appendix:1}

In this appendix, we provide detailed explanations on the feature engineering process used in our study, focusing on how each feature is retrieved and calculated from fields in the issue reports within JIRA.

\subsection*{JIRA Issue Report Fields}
\begin{figure}[!htbp]
  \centering
  \includegraphics[width=\textwidth]{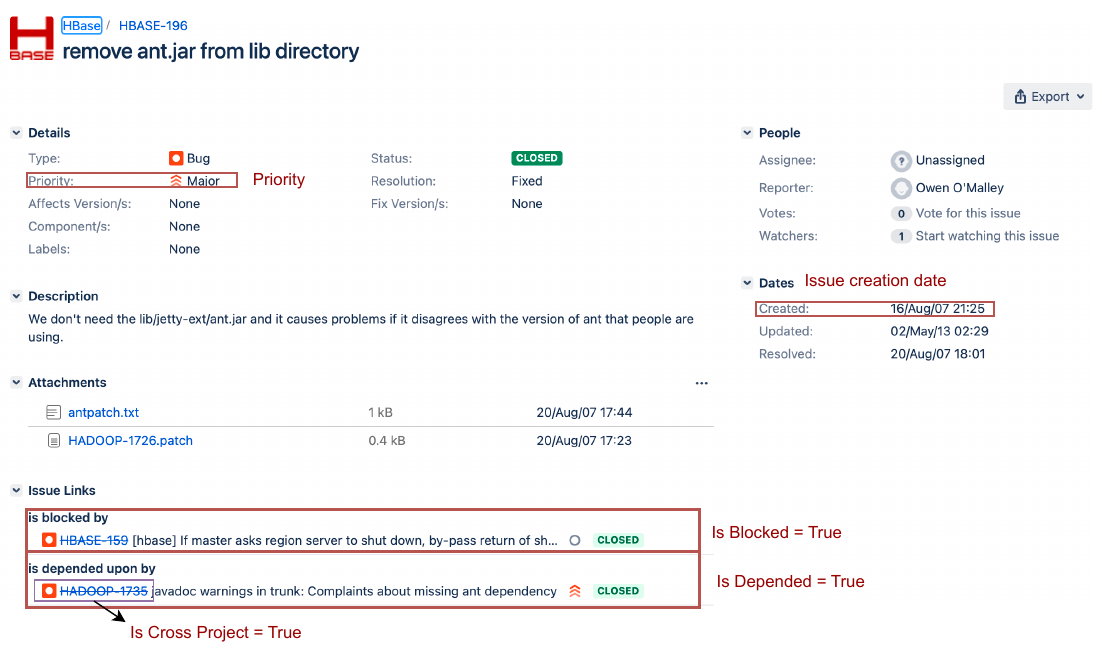}
  \caption{Example JIRA Issue Report showing how \texttt{Priority}, \texttt{Is Blocker} and \texttt{Is Dependened} and \textbf{Number of Parallel Issues} are extracted}
  \label{fig:jira_example}
\end{figure}

\begin{itemize}
    \item \textbf{Priority}, \textbf{Is Duplicate}, \textbf{Is Blocker}, \textbf{Is Blocked}, \textbf{Is Regression}, \textbf{Is Dependent}, \textbf{Is Incorporates}, \textbf{Isli Required}, \textbf{Is Reference}, \textbf{Is Completes}, \textbf{Is Testing}, \textbf{Is Issue Split}, \textbf{Is Supercedes}, \textbf{Is Cloner}, \textbf{Is Container}, \textbf{Is Parent Feature}, and \textbf{Is Child-Issue} can all be directly retrieved from the corresponding fields in the JIRA issue report. Figure~\ref{fig:jira_example} illustrates how features such as \textbf{Priority}, \textbf{Is Blocker}, and \textbf{Is Dependent} are extracted from a sample issue report. Note that not all link-type relationships are visible in an issue’s report page—if an issue has no recorded relationship with other issues, those corresponding features are treated as \textbf{False} by default.

    \item \textbf{Number of Parallel Issues}, for each issue report, we calculate the number of issues that were opened on the same day, based on the report’s creation date, and define this as the number of parallel opened issues.

    \item \textbf{Is Cross Projects}: As shown in Figure~\ref{fig:jira_example}, this feature is set to \textbf{True} when the issue report contains a link to another issue that belongs to a different module or project, as indicated in the "Link" field.    

    \begin{figure}[!htbp]
      \centering
      \includegraphics[width=0.7\textwidth]{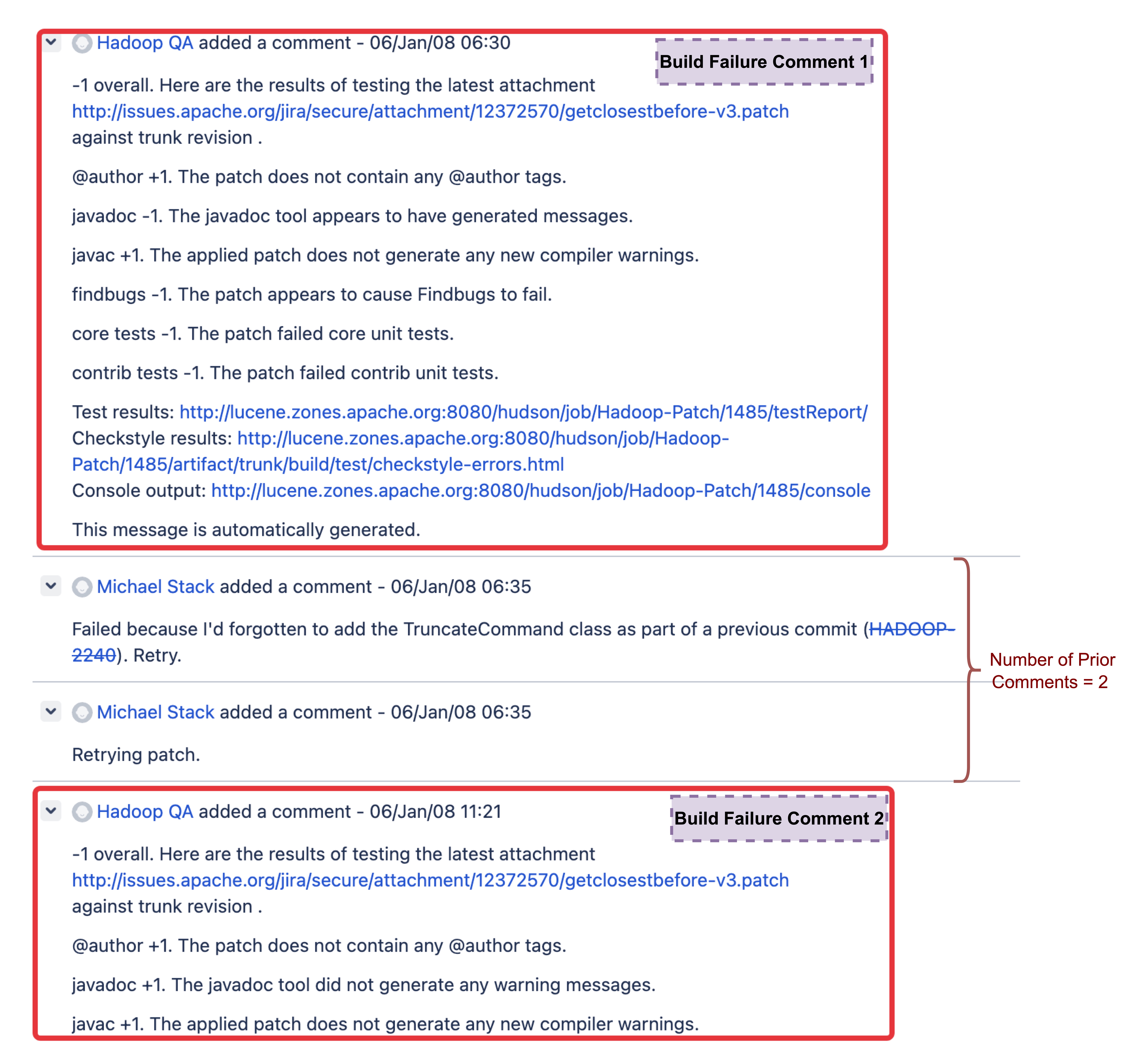}
      \caption{Example of Calculating the Number of Prior Comments}
      \label{fig:jira_example_2}
    \end{figure}
    
    \item \textbf{Number of Prior Comments}: As shown in Figure~\ref{fig:jira_example_2}, this feature represents the number of comments made before the comment reporting the build failure. For example, in the case of Build Failure 2, there are two comments preceding the failure comment, so the number of prior comments is 2.

    \begin{figure}[!htbp]
      \centering
      \includegraphics[width=0.7\textwidth]{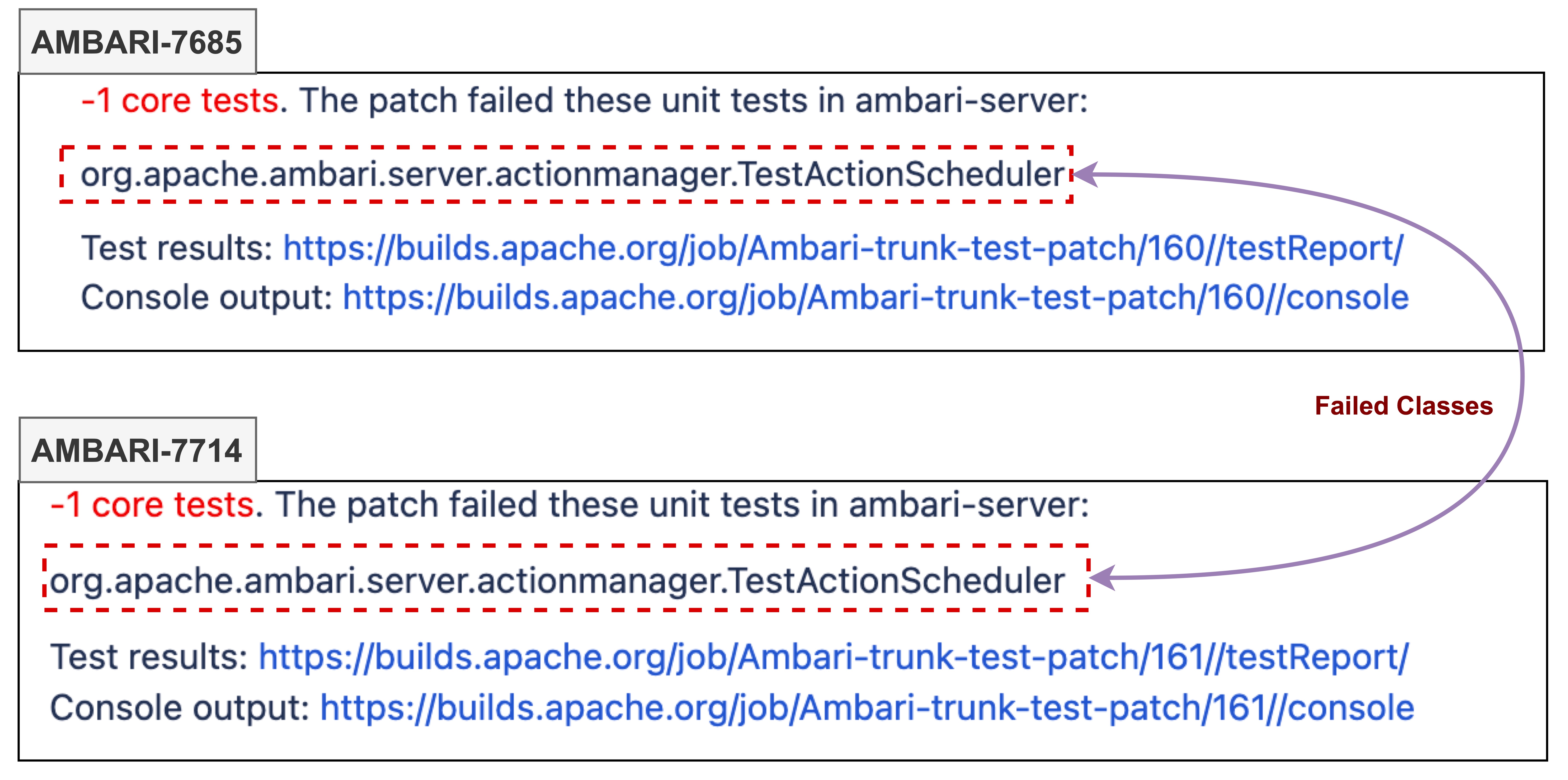}
      \caption{Example of Retrieving the Failed Classes from Build Logs}
      \label{fig:jira_example_3}
    \end{figure}
    
   \item \textbf{Is Shared Same Emsg} and \textbf{Number of Similar Failures}: These two features are derived by extracting the failed classes from the failure logs. As illustrated in Figure~\ref{fig:jira_example_3}, the failed classes are parsed from the build failure comments. In this example, both failure comments report the same error message (Emsg), resulting in \textbf{Is Shared Same Emsg} being set to \textbf{True}, and the \textbf{Number of Similar Failures} being equal to 1.

    \begin{figure}[!htbp]
      \centering
      \includegraphics[width=0.7\textwidth]{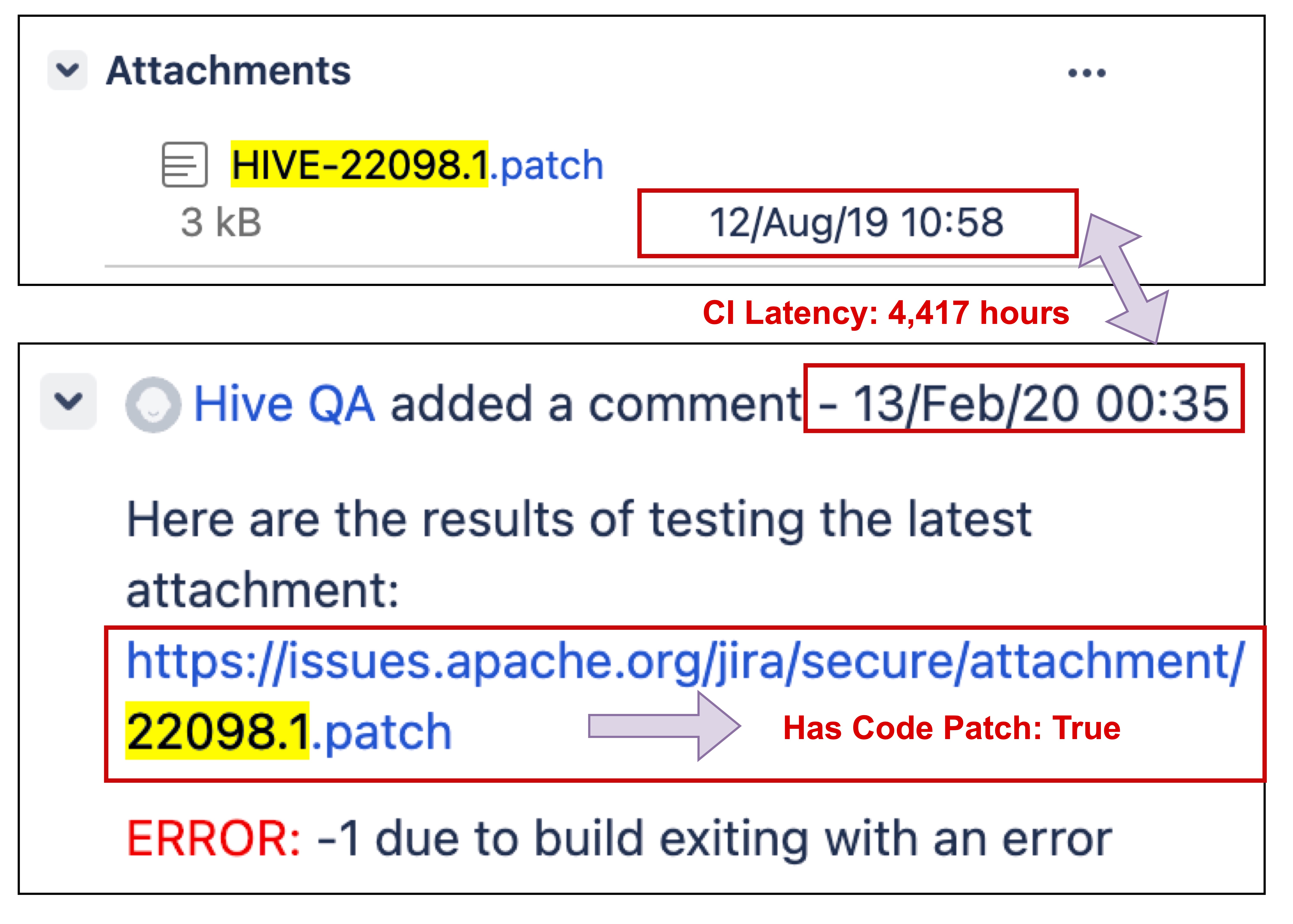}
      \caption{Example of Retrieving the Has Code Patch and CI Latency}
      \label{fig:jira_example_4}
    \end{figure}
    
    \item \textbf{CI Latency}: This metric is calculated by taking the difference between the timestamp when the code patch is attached to the JIRA issue and the timestamp when the CI build is triggered. Figure~\ref{fig:jira_example_4} illustrates how these timestamps are retrieved and how the CI latency is computed in the given example.

    \item \textbf{Has Code Patch}: This feature is determined by checking whether the build failure comment includes an attached code patch. Figure~\ref{fig:jira_example_4} shows where the code patch appears in the JIRA issue report.    
    
\end{itemize}